\documentclass[12pt,draftcls,peerreview,onecolumn]{IEEEtran}

\usepackage{theorem}
\usepackage{times}
\usepackage[dvips]{graphicx,color}
\usepackage{exscale}
\usepackage{epic,eepic}
\usepackage{amsmath}
\usepackage{amssymb}
\usepackage{cite, verbatim}
\usepackage{psfrag}

\newtheorem{definition}{Definition}
\newtheorem{theorem}{Theorem}
\newtheorem{lemma}{Lemma}

\newcommand{\qed}{\nobreak \ifvmode \relax \else
      \ifdim\lastskip<1.5em \hskip-\lastskip
      \hskip1.5em plus0em minus0.5em \fi \nobreak
      \vrule height0.75em width0.5em depth0.25em\fi}

\begin{document}

\title{Capacity Definitions for General Channels with Receiver Side Information}
\author{Michelle~Effros,~\IEEEmembership{Senior Member,~IEEE,}
        Andrea~Goldsmith,~\IEEEmembership{Fellow,~IEEE,}
        and~Yifan~Liang,~\IEEEmembership{Student~Member,~IEEE}
\thanks{
This work was supported by the DARPA ITMANET program under grant
number 1105741-1-TFIND. The material in this paper was presented in
part at the IEEE International Symposium on Information Theory,
Cambridge, Massachusetts, August 1998 and the IEEE International
Symposium on Information Theory, Nice, France, June 2007.
}
\thanks{M. Effros is with the Department
of Electrical Engineering, California Institute of Technology,
Pasadena, CA 91125 (email: effros@caltech.edu).}
\thanks{A. Goldsmith and Y. Liang are with the Department of~Electrical Engineering, Stanford
University, Stanford CA 94305 (email: andrea@wsl.stanford.edu;
yfl@wsl.stanford.edu).}
}



%

\maketitle

\newcommand{\beq}{\begin{equation}}
\newcommand{\eeq}{\end{equation}}
\newcommand{\ubar}{\underline}
\newcommand{\typ}{{A_\epsilon^{(n)}}}
\newcommand{\Pens}{P_e^{(n,s)}}
\newcommand{\bs}[1]{\boldsymbol{#1}}
\newcommand{\Fig}[2]
{\begin{figure}[htbp]
\begin{center}
\includegraphics[width=3in]{EPS/#1.eps}
\caption{#2} \label{fig:#1}
\end{center}
\end{figure}}
\newcommand{\refE}[1] {Eqn.~(\ref{eqn:#1})}
\newcommand{\refS}[1] {Section~\ref{sec:#1}}
\newcommand{\refF}[1] {Fig.~\ref{fig:#1}}
\newcommand{\refA}[1] {Appendix \ref{app:#1}}
\newcommand{\bsi}[1] {b_s^{(#1)}}
\newcommand{\bpi}[1] {b_p^{(#1)}}
\newcommand{\calIn}[1]{\mathcal{I}_{n,#1}}
\newcommand{\BC} {\textrm{BC}}

\begin{abstract}
We consider three capacity definitions for general channels with
channel side information at the receiver, where the channel is
modeled as a sequence of finite dimensional conditional
distributions not necessarily stationary, ergodic, or information
stable. The {\em Shannon capacity} is the highest rate
asymptotically achievable with arbitrarily small error probability.
The {\em capacity versus outage} is the highest rate asymptotically
achievable with a given probability of decoder-recognized outage.
The {\em expected capacity} is the highest average rate
asymptotically achievable with a single encoder and multiple
decoders, where the channel side information determines the decoder
in use. As a special case of channel codes for expected rate, the
code for capacity versus outage has two decoders: one operates in
the non-outage states and decodes all transmitted information, and
the other operates in the outage states and decodes nothing.
Expected capacity equals Shannon capacity for channels governed by a
stationary ergodic random process but is typically greater for
general channels. These alternative capacity definitions essentially
relax the constraint that all transmitted information must be
decoded at the receiver. We derive capacity theorems for these
capacity definitions through information density. Numerical examples
are provided to demonstrate their connections and differences. We
also discuss the implication of these alternative capacity
definitions for end-to-end distortion, source-channel coding and
separation.
\end{abstract}

\pagebreak

\begin{IEEEkeywords}
Composite channel, Shannon capacity, capacity versus outage, outage
capacity, expected capacity, information density, broadcast
strategy, binary symmetric channel (BSC), binary erasure channel
(BEC), source-channel coding, separation.
\end{IEEEkeywords}

\IEEEpeerreviewmaketitle

\section{Introduction}
Channel capacity has a natural {\em operational definition}: the
highest rate at which information can be sent with arbitrarily low
probability of error \cite[p. 184]{CoverIT}. Channel coding
theorems, a fundamental subject of Shannon theory, focus on finding
{\em information theoretical definitions} of channel capacity, i.e.
expressions for channel capacity in terms of the probabilistic
description of various channel models.

In his landmark paper \cite{Shannon48}, Shannon showed the capacity
formula
\begin{equation}
C = \max_X I(X;Y) \label{eqn:ShannonCap}
\end{equation}
for {\em memoryless} channels. The capacity formula
\eqref{eqn:ShannonCap} is further extended to the well-known
limiting expression
\begin{equation}
C = \lim_{n\to \infty} \sup_{X^n} \frac1n I(X^n;Y^n)
\label{eqn:infoStableCap}
\end{equation}
for channels with memory. Dobrushin proved the capacity formula
\eqref{eqn:infoStableCap} for the class of {\em information stable}
channels in \cite{Dobrushin63}. However, there are channels that do
not satisfy the information stable condition and for which the
capacity formula \eqref{eqn:infoStableCap} fails to hold. Examples
of information unstable channels include the stationary regular
decomposable channels \cite{Winkelbauer71}, the stationary
nonanticipatory channels \cite{Kieffer74} and the averaged
memoryless channels \cite{Ahlswede68}. In \cite{VerduH:94} Verd\'u
and Han derived the capacity
\begin{equation}
 C = \sup_{\bs{X}} \underline{\bs{I}} (\bs{X}; \bs{Y})
 \label{eqn:VerduHanC}
\end{equation}
for general channels, where $\underline{\bs{I}} (\bs{X}; \bs{Y})$ is
the liminf in probability of the normalized information density. The
completely general formula \eqref{eqn:VerduHanC} does not require
any assumption such as memorylessness, information stability,
stationarity, causality, etc.

The focus of this paper is on one class of such information unstable
channels, the {\em composite channel} \cite{Effros98}. A composite
channel is a collection of channels $\{W_s:s\in{\cal S}\}$
parameterized by $s$, where each component channel is stationary and
ergodic. The channel realization is determined by the random
variable $S$, which is chosen according to some channel state
distribution $p(s)$ at the beginning of transmission and then held
fixed. The composite channel model describes many communication
systems of practical interest, for instance, applications with
stringent delay constraint such that a codeword may not experience
all possible channel states, systems with receiver complexity
constraint such that decoding over long blocklength is prohibited,
and slow fading wireless channels with channel coherence time longer
than the codeword duration. Ahlswede studied this class of channels
under the name {\em averaged channel} and obtained a formula for
Shannon capacity in \cite{Ahlswede68}. It is also referred to as the
{\em mixed channel} in \cite{HanBook}. The class of composite
channels can be generalized to channels for which the optimal input
distribution induces a joint input-output distribution on which the
ergodic decomposition theorem \cite[Theorem 1.8.2]{Gray90} holds,
e.g. stationary distributions defined on complete, separable metric
spaces (Polish spaces). In this case the channel index $s$ becomes
the ergodic mode.

Shannon's capacity definition, with a focus on stationary and
ergodic channels, has enabled great insight and design inspiration.
However, the definition is based on asymptotically large delay and
imposes the constraint that all transmitted information be correctly
decoded. In the case of composite channels the capacity is dominated
by the performance of the ``worst'' component channel, no matter how
small its probability. This highlights the pessimistic nature of the
Shannon capacity definition, which forces the use of a single code
with arbitrarily small error probability. In generalizing the
channel model to deal with such scenarios as the composite channel
above, we relax the constraints and generalize the capacity
definitions. These new definitions are fundamental, and they address
practical design strategies that give better performance than
traditional capacity definitions.

Throughout this paper we assume the channel state information is
revealed to the receiver (CSIR), but no channel state information is
available at the transmitter (CSIT). The downlink satellite
communication system gives an example where the transmitter may not
have access to CSIT: the terrestrial receivers implement channel
estimation but do not have sufficient transmit power to feed back
the channel knowledge to the satellite transmitter. In other cases,
the transmitter may opt for simplified strategies which do not
implement any adaptive transmission based on channel state, and
therefore CSIT becomes irrelevant.

The first alternative definition we consider is {\em capacity versus
outage} \cite{OzarowS:94}. In the absence of CSIT, the transmitter
is forced to use a single code, but the decoder may decide whether
the information can be reliably decoded based on CSIR. We therefore
design a coding scheme that works well most of the time, but with
some maximal probability $q$, the decoder sees a bad channel and
declares an outage; in this case, the transmitted information is
lost. The encoding scheme is designed to maximize the capacity for
non-outage states. Capacity versus outage was previously examined in
\cite{OzarowS:94} for single-antenna cellular systems, and later
became a common criterion used in multiple-antenna wireless fading
channels \cite{FoschiniG:97, MIMOJSAC, Zheng02}. In this work we
formalize the operational definition of capacity versus outage and
also give the information-theoretical definition through the
distribution of the normalized information density.

Another method for dealing with channels of variable quality is to
allow the receiver to decode {\em partial} transmitted information.
This idea can be illustrated using the broadcast strategy suggested
by Cover~\cite{Cover72}. The transmitter views the composite channel
as a broadcast channel with a collection of virtual receivers
indexed by channel realization $S$. The encoder uses a broadcast
code and encodes information as if it were broadcasting to the
virtual receivers. The receiver 
chooses the appropriate decoder for the broadcast code based on the
channel $W_S$ in action. The goal is to identify the point in the
broadcast rate region that maximizes the {\em expected rate}, where
the expectation is taken with respect to the state distribution
$p(S)$ on ${\cal S}$. Shamai et al. first derived the expected
capacity for Gaussian slowly fading channels in \cite{Shamai97} and
later extended the result to MIMO fading channels in
\cite{Shamai03}. The formal definition of expected capacity was
introduced in \cite{Effros98}, where upper and lower bounds were
also derived for the expected capacity of any composite channel.
Details of the proofs together with a numerical example of a
composite binary symmetric channel (BSC) appeared recently in
\cite{Liang072}. Application of the broadcast strategy to minimize
the end-to-end expected distortion is also considered in
\cite{Gunduz,Ng07b}.

The alternative capacity definitions are of particular interest for
applications where it is desirable to maximize average received rate
even if it means that part of the transmitted information is lost
and the encoder does not know the exact delivered rate. In this case
the receiver either tolerates the information loss or has a
mechanism to recover the lost information. Examples include
scenarios with some acceptable outage probability, communication
systems using multiresolution or multiple description source codes
such that partial received information leads to a coarse but still
useful source reconstruction at a larger distortion level, feedback
channels where the receiver tells the transmitter which symbols to
resend, or applications where lost source symbols are well
approximated by surrounding samples. The received rate averaged over
multiple transmissions is a meaningful metric when there are two
time horizons involved: a short time horizon at the end of which
decoding has to be performed because of stringent delay constraint
or decoder complexity constraint, and a long time horizon at the end
of which the overall throughput is evaluated. For example, consider
a wireless LAN service subscriber. Whenever the user requests a
voice or data transmission over the network, he usually expects the
information to be delivered within a couple of minutes, i.e. the
short time horizon. However, the service charge is typically
calculated on a monthly basis depending on the total or average
throughput within the entire period, i.e. the long time horizon.

It is worth pointing out that our capacity analysis does not apply
to the {\em compound channel}
\cite{CsiszarK:81,BlackwellB:59,Wolfowitz:64}. A compound channel
includes a collection of channels but does not assume any associated
state distribution and therefore has no information density
distribution, on which the capacity definition relies. Our channel
model also excludes the {\em arbitrarily varying channel}
\cite{CsiszarK:81,CsiszarN:91}, where the channel state changes on
each transmission in a manner that depends on the channel input in
order to minimize the capacity of the chosen encoding and decoding
strategies.

The remainder of this paper is structured as follows. In Section
\ref{sec:defn} we review how the information theoretical definitions
of channel capacity evolved with channel models, and give a few
definitions that serve as the basis for the development of
generalized capacity definitions. The Shannon capacity is considered
in Section \ref{sec:Shannon}, where we provide an alternative proof
of achievability based on a modified notion of typical sets. We also
show that the Shannon capacity only depends on the support set of
the channel state distribution. In Section \ref{sec:outageCapacity}
we give a formal definition of the capacity versus outage and
compare it with the closely-related concept of $\epsilon$-capacity
\cite{VerduH:94}. In Section \ref{sec:expectedCap} we introduce the
expected capacity and establish a bijection between the
expected-rate code and the broadcast channel code. In Section
\ref{sec:ex} we compare capacity definitions and their implications
through two examples: the Gilbert-Elliott channel and the BSC with
random crossover probabilities. The implication of these alternative
capacity definitions for end-to-end distortion, source-channel
coding and separation is briefly discussed in Section
\ref{sec:sccoding}. Conclusions are given in Section \ref{sec:con}.

\section{Background} \label{sec:defn}
Shannon in \cite{Shannon48} defined the channel capacity as the
supremum of all achievable rates $R$ for which there exists a
sequence of $(2^{nR}, n)$ codes such that the probability of error
tends to zero as the blocklength $n$ approaches infinity, and showed
the capacity formula \eqref{eqn:ShannonCap}
\[
C = \max_X I(X;Y)
\]
for {\em memoryless} channels. In proving the capacity formula
\eqref{eqn:ShannonCap}, the converse of the coding theorem \cite[p.
206]{CoverIT} uses Fano's inequality and establishes the right-hand
side of \eqref{eqn:ShannonCap} as an upper bound of the rate of any
sequence of channel codes with error probability approaching zero.
The direct part of the coding theorem then shows any rate below the
capacity is indeed achievable. Although the capacity formula
\eqref{eqn:ShannonCap} is a single-letter expression, the direct
channel coding theorem requires coding over long blocklength
to achieve arbitrarily small error probability. The receiver decodes
by joint typicality with the typical set defined as \cite[pp.
195]{CoverIT}
\begin{eqnarray}
A_\epsilon^{(n)} = && \Big\{ (x^n, y^n) \in \mathcal{X}^n \times
\mathcal{Y}^n:  \nonumber \\
&& \left|-\frac{1}{n}\log p(x^n)
- H(X)\right| < \epsilon, \nonumber \\
&& \left|-\frac{1}{n}\log p(y^n) - H(Y)\right| <
\epsilon, \nonumber \\
&& \left. \left|-\frac{1}{n}\log p(x^n, y^n) - H(X, Y)\right| <
\epsilon \right\}, \label{eqn:jointTypicalSet}
\end{eqnarray}
which relies on the law of large numbers to obtain the asymptotic
equipartition property (AEP).

For channels with memory, the capacity formula
\eqref{eqn:ShannonCap} generalizes to the limiting expression
\eqref{eqn:infoStableCap}
\[
C = \lim_{n\to \infty} \sup_{X^n} \frac1n I(X^n;Y^n).
\]
However, the capacity formula \eqref{eqn:infoStableCap} does not
hold in full generality. Dobrushin proved it for the class of
information stable channels. The class of information stable
channels, including the class of memoryless channels as a special
case, can be roughly described as having the property that the input
maximizing the mutual information $I(X^n;Y^n)$ and its corresponding
output behave ergodically. In a sense, an ergodic sequence is the
most general dependent sequence for which the strong law of large
numbers holds \cite[p. 474]{CoverIT}. The coding theorem of
information stable channels follows similarly from that of
memoryless channels.

However, the joint typicality decoding technique cannot be
generalized to information unstable channels. For general channels,
the set $A_\epsilon^{(n)}$ defined in \eqref{eqn:jointTypicalSet}
does not have the AEP. As an evidence, the probability of
$A_\epsilon^{(n)}$ does not approach $1$ for large $n$. We may not
construct channel codes which has small error probability and
meanwhile has a rate arbitrarily close to 
\eqref{eqn:infoStableCap}.
Therefore,
the right-hand side of \eqref{eqn:infoStableCap}, although still a
valid upper bound of channel capacity, is not necessarily tight. In
\cite{VerduH:94} Verd\'u and Han presented a tight upper bound for
general channels and showed its achievability through Feinstein's
lemma \cite{Feinstein54}. We provide an alternative proof of
achievability based on a new notion of typical sets in Section
\ref{sec:Shannon}.

This information stable condition can be illustrated using the
concept of {\em information density}.
\begin{definition}[Information Density]
\label{defn:infoDensity} {\em Given a joint distribution
$P_{X^nY^n}$ on ${\cal X}^n \times {\cal Y}^n$ with marginal
distributions $P_{X^n}$ and $P_{Y^n}$, the} information density {\em
is defined as \cite{Han93}}
\begin{eqnarray}
i_{X^nY^n}(x^n;y^n) &=& \log\frac{P_{X^nY^n}(x^n,
y^n)}{P_{X^n}(x^n)P_{Y^n}(y^n)}
\nonumber \\
&=& \log\frac{P_{Y^n|X^n}(y^n|x^n)}{P_{Y^n}(y^n)}.
\label{eqn:infoden1}
\end{eqnarray}
\end{definition}

\noindent The distribution of the random variable
$(1/n)i_{X^nY^n}(x^n;y^n)$ is referred to as the {\em information
spectrum} of $P_{X^nY^n}$. It is observed that the normalized mutual
information
\[
\frac1n I(X^n; Y^n) = \sum_{(x^n, y^n)} p(x^n, y^n) \cdot \frac1n
\log \frac{p(y^n|x^n)}{p(y^n)}
\]
is the expectation of the normalized information density
\[
\frac1n i(x^n; y^n) = \frac1n \log \frac{p(y^n|x^n)}{p(y^n)}
\]
with respect to the underlying joint input-output distribution
$p(x^n, y^n)$, i.e.
\[
\frac1n I(X^n;Y^n) = \mathbb{E}_{X^nY^n} \left\{ \frac1n
i_{X^nY^n}(X^n;Y^n)\right\}.
\]
Denote by $X^n_*$ the input distribution that maximizes the mutual
information $I(X^n;Y^n)$ and by $Y^n_*$ the corresponding output
distribution. The information stable condition \cite[Definition
3]{Vembu95} requires that the normalized information density
$(1/n)i(X^n_*; Y^n_*)$, as a random variable, converges in
distribution to a constant equal to the normalized mutual
information $(1/n)I(X^n_*; Y^n_*)$ as the blocklength $n$ approaches
infinity.

In \cite{VerduH:94} Verd\'u and Han derived the capacity formula
\eqref{eqn:VerduHanC}
\[
C = \sup_{\bs{X}} \underline{\bs{I}} (\bs{X}; \bs{Y})
\]
for general channels, where $\underline{\bs{I}} (\bs{X}; \bs{Y})$ is
the liminf in probability of the normalized information density. In
contrast to information stable channels where the distribution of
$(1/n)i(X^n; Y^n)$ converges to a single point, for information
unstable channels, even with infinite blocklength the {\em support
set}\footnote{The smallest closed set of which the complement set
has probability measure zero.}
of the distribution of $(1/n)i(X^n; Y^n)$ may still have multiple
points or even contain an interval. The Shannon capacity equals the
infimum of this support set.

The information spectrum of an information stable channel is
demonstrated in the upper plot of \refF{converge}. As the block
length $n$ increases, the convergence of the normalized information
density to the channel capacity follows from the weak law of large
numbers. In the lower plot of \refF{converge}, we show the empirical
distribution of $(1/n)i(X^n;Y^n)$ for an information unstable
channel. The distribution of the normalized information density does
not converge to a single point, so the equation
\eqref{eqn:infoStableCap} does not equal the capacity, which is
given by \eqref{eqn:VerduHanC}.
\psfrag{p1nixy}{$p\left(\frac1ni(x^n;y^n)\right)$}
\psfrag{Iunderbar}{$\underline{\bs{I}} (\bs{X}; \bs{Y})$}
\psfrag{1nixnyn}{$\frac1ni(x^n;y^n)$}
\psfrag{lim1nIxnyn}{${\displaystyle \lim_{n\to\infty}}
\frac1nI(X^n;Y^n)$} \psfrag{n=1}{$n=1$} \psfrag{n=10}{$n=10$}
\psfrag{n=1000}{$n=1000$} \psfrag{n=infty}{$n=\infty$}
\begin{figure}[htbp]
\begin{center}
\includegraphics[width=3in]{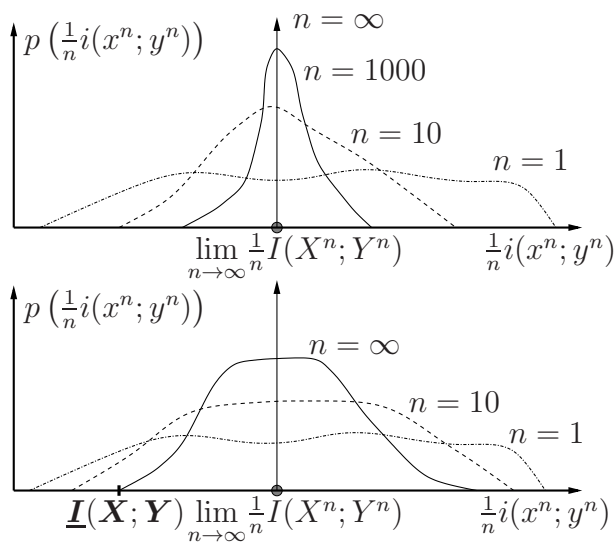}
\caption{Empirical distribution of normalized information density.
Upper: information stable channel. Lower: information unstable
channel.} \label{fig:converge}
\end{center}
\end{figure}

\section{Shannon Capacity}
\label{sec:Shannon} We consider a channel $\bs{W}$ which is
statistically modeled as a sequence of $n$-dimensional conditional
distributions $\bs{W}=\{W^n=P_{Z^n|X^n}\}_{n=1}^\infty$. For any
integer $n>0$, $W^n$ is the conditional distribution from the input
space ${\cal X}^n$ to the output space ${\cal Z}^n$. Let $\bs{X}$
and $\bs{Z}$ denote the input and output processes, respectively,
for the given sequence of channels. Each process is specified by a
sequence of finite-dimensional distributions, e.g.
$\bs{X}=\{X^n=(X_1^{(n)},\cdots,X_n^{(n)})\}_{n=1}^\infty$.

To consider the special case where the decoder has receiver side
information not present at the encoder, we represent this side
information as an additional output of the channel. Specifically, we
let $Z^n=(S,Y^n)$, where $S$ is the channel side information and
$Y^n$ is the output of the channel described by parameter $S$.
Throughout, we assume that $S$ is a random variable independent of
$\bs{X}$ and unknown to the encoder. Thus for each $n$
\[
P_{W^n}(z^n|x^n)= P_{Z^n|X^n}(s, y^n|x^n)  = P_{S}(s)
P_{Y^n|X^n,S}(y^n|x^n,s),
\]
and the information density \eqref{eqn:infoden1} can be rewritten as
\begin{eqnarray}
i_{X^nW^n}(x^n;z^n)
&=& \log\frac{P_{W^n}(z^n|x^n)}{P_{Z^n}(z^n)} \nonumber \\
&=& \log\frac{P_{Y^n|X^n,S}(y^n|x^n,s)}{P_{Y^n|S}(y^n|s)} \nonumber \\
&=& i_{X^nW^n}(x^n;y^n|s). \label{eqn:infoden}
\end{eqnarray}
In the following we see that the generalized capacity definitions of
composite channels depend crucially on information density instead
of mutual information. We also denote by $F_{\bs{X}}(\alpha)$ the
limit of the cumulative distribution function (cdf) of the
normalized information density, i.e.
\begin{equation}
F_{\bs{X}}(\alpha) = \lim_{n \to \infty}
P_{X^nW^n} \left\{ \frac1n i_{X^nW^n}(X^n;Y^n|S) \le \alpha
\right\}, \label{eqn:cdfFalpha}
\end{equation}
where the subscript emphasizes the input process $\bs{X}$.

Consider a sequence of $(2^{nR},n)$ codes for channel $\bs{W}$,
where for any $R > 0$, a $(2^{nR},n)$ code is a collection of
$2^{nR}$ blocklength-$n$ channel codewords and the associated
decoding regions.  The Shannon capacity is defined as the supremum
of all rates $R$ for which there exists a sequence of $(2^{nR}, n)$
codes with vanishing error probability \cite{Shannon48}. Therefore,
the Shannon capacity $C(\bs{W})$ measures the rate that can be
reliably transmitted from the encoder and also be reliably received
at the decoder. We simplify this notation to $C$ if the channel
argument is clear from context.

The achievability and converse theorems for the Shannon capacity of
a general channel
\begin{eqnarray}
C &=& \sup_{\bs{X}} \ubar{\bs{I}}(\bs{X};\bs{Z}) =  \sup_{\bs{X}}
\ubar{\bs{I}}(\bs{X};\bs{Y}|S)
\nonumber \\
&=& \sup_{\bs{X}} \sup \left\{ \alpha: F_{\bs{X}}(\alpha) = 0
\right\} \label{eqn:ShannonCapFormula}
\end{eqnarray}
are proved, respectively, by Theorems $2$ and $5$ of
\cite{VerduH:94}, using Feinstein's lemma \cite{Feinstein54},
\cite[Lemma 3.4.1]{HanBook}, \cite[Lemma 3.5.2]{AshBook} and the
Verd\'{u}-Han lemma \cite[Theorem 4]{VerduH:94}. The special case of
a composite channel with CSIR follows immediately from this result.
We here provide an alternative proof of achievability based on a
modified notion of typical sets. In the following proof we simplify
notations by removing the explicit conditioning on the side
information $S$.

{\em Encoding}: For any input distribution $P_{X^n}$, $\epsilon >
0$, and $R < \ubar{\bs{I}}(\bs{X};\bs{Y})-\epsilon$, generate the
codebook by choosing $X^n(1)$, $\cdots$, $X^n(2^{nR})$ i.i.d.
according to the distribution $P_{X^n}(x^n)$.

{\em Decoding}: For any $\epsilon>0$, the {\em typical set} $\typ$
is defined as
\begin{equation}
\typ = \left\{ (x^n,y^n):\frac1ni_{X^nW^n}(x^n;y^n)\geq
\ubar{\bs{I}}(\bs{X};\bs{Y})-\epsilon \right\}. \label{eqn:typ}
\end{equation}
Channel output $Y^n$ is decoded to $X^n(i)$ where $i$
is the unique index for which
$(X^n(i),Y^n)\in\typ$. An error is declared if more than one or no
such index exists.

{\em Error Analysis}: We define the following events for all indices
$1 \le i, j \le 2^{nR}$,
\begin{equation}
E_{ji} = \left\{ \left. (X^n(j),Y^n) \in \typ  \right| X^n(i) \,\,
\text{sent} \right\}. \label{eqn:EventEij}
\end{equation}
Conditioned on codeword $X^n(i)$ being sent, the probability of the
corresponding error event $E_i$
\[
E_i = \bigcup_{j \neq i} E_{ji} \bigcup E_{ii}^c,
\]
can be bounded by
\[
\Pr(E_i) \le \Pr(E_{ii}^c) + \sum_{j \neq i} \Pr(E_{ji}).
\]
Since we generate i.i.d. codewords, $\Pr(E_{ii})$ and $\Pr(E_{ji})$,
$j\neq i$, do not depend on the specific indices $i$, $j$. Assuming
equiprobable inputs, the expected probability of error with respect
to the randomly generated codebook is:
\begin{eqnarray}
&& P_e^{(n)} \nonumber \\
&=& \Pr \left\{ \mbox{error}|X^n(1)\mbox{ sent} \right\}
\nonumber \\
&\le& \Pr(E_{11}^c) + \sum_{j=2}^{2^{nR}} \Pr(E_{j1}) \nonumber \\
& \le & P_{X^nW^n} \left[ \frac1n
i_{X^nW^n}(X^n(1);Y^n)<\ubar{\bs{I}}(\bs{X};\bs{Y})-\epsilon \right] \nonumber  \\
&& + \,\, 2^{nR}\sum_{(x^n,y^n)\in\typ} P _{X^n} (x^n)P _{Y^n}
(y^n) \nonumber \\
& \le & \epsilon_n + 2^{n[R-\ubar{\bs{I}}(\bs{X};\bs{Y})+\epsilon]}
        \sum_{(x^n,y^n)\in\typ} P_{X^nW^n}(x^n,y^n), 
        \label{eqn:tradcap}
\end{eqnarray}
where by definition of $\ubar{\bs{I}}(\bs{X};\bs{Y})$ we have
$\epsilon_n$ approaching $0$ for $n$ large enough. The last
inequality uses \eqref{eqn:infoden}, \eqref{eqn:typ}, and the fact
that $(x^n,y^n)\in\typ$ implies
\[
\frac1n i_{X^nW^n}(x^n;y^n) = \frac1n
\log\frac{P_{X^nW^n}(x^n,y^n)}{P_{X^n}(x^n)P_{Y^n}(y^n)} \ge
\ubar{\bs{I}}(\bs{X};\bs{Y}) -\epsilon
\]
and consequently
\[
P_{X^n}(x^n)P_{Y^n}(y^n)  \le
2^{-n[\ubar{\bs{I}}(\bs{X};\bs{Y})-\epsilon]}P_{X^nW^n}(x^n, y^n).
\]
From \eqref{eqn:tradcap}
\[
P_e^{(n)} \leq \epsilon_n +
2^{n[R-\ubar{\bs{I}}(\bs{X};\bs{Y})+\epsilon]}\rightarrow 0
\]
for all $R < \ubar{\bs{I}}(\bs{X};\bs{Y})-\epsilon$ and arbitrary
$\epsilon>0$, which completes our proof.

Although a composite channel is characterized by the collection of
component channels $\{W_s: s \in \mathcal{S}\}$ and the associated
probability distribution $p(s)$ on $\mathcal{S}$, the Shannon
capacity of a composite channel is solely determined by the support
set of the channel state distribution $p(s)$. In the case of a
discrete channel state set $\mathcal{S}$, we only need to know which
channel states have positive probability. The exact positive value
that the probability mass function $p(s)$ assigns to channel states
is irrelevant in view of the Shannon capacity. In the case of a
continuous channel state set $\mathcal{S}$, we only need to know the
subset of channel states where the probability density function is
strictly positive. This is formalized in Lemma
\ref{lemma:sameShannonC}. Before introducing the lemma we need the
following definition \cite[Appendix 8]{Durrett05}.
\begin{definition}[Equivalent Probability Measure]
{\em A probability measure $p_1$ is} absolutely continuous with
respect to {\em $p_2$, written as $p_1 \ll p_2$, if $p_1(A) = 0$
implies that $p_2(A)=0$ for any event $A$. Here $p_i(A)$, $i=1,2$,
is the probability of event $A$ under probability measure $p_i$.
$p_1$ and $p_2$ are} equivalent probability measures {\em if $p_1
\ll p_2$ and $p_2 \ll p_1$.}
\end{definition}
\begin{lemma}\label{lemma:sameShannonC}
{\em Consider two composite channels $\bs{W}_1$ and $\bs{W}_2$ with
component channels from the same collection $\{W_s: s \in
\mathcal{S}\}$. Denote by $p_1(s)$ and $p_2(s)$, respectively, the
corresponding channel state distribution of each composite channel.
Then $p_1 \ll p_2$ implies $C(\bs{W}_1) \le C(\bs{W}_2)$.
Furthermore, if $p_1$ and $p_2$ are equivalent probability measures,
then  $C(\bs{W}_1) = C(\bs{W}_2)$.}
\end{lemma}
Intuitively speaking, $p_1 \ll p_2$ if the support set for
$\bs{W}_2$ is a subset of the support set for $\bs{W}_1$, so any
input distribution that allows reliable transmission on $\bs{W}_1$
also allows reliable transmission on $\bs{W}_2$. $p_1$ and $p_2$ are
equivalent probability measures if they share the same support set,
and this guarantees that the corresponding composite channels have
the same Shannon capacity. Details of the proof are given in
\refA{ShannonC}.

The equivalent probability measure is a sufficient but not necessary
condition for two composite channels to have the same Shannon
capacity. For example, consider two slow-fading Gaussian composite
channels. It is possible that two probability measures have no
support below the same channel gain, but one assigns non-zero
probability to states with large capacity while the other does not.
In this case, the probability measures are not equivalent;
nevertheless the Shannon capacity of both composite channels are the
same.

\section{Capacity versus Outage}
\label{sec:outageCapacity} The Shannon capacity definition imposes
the constraint that all transmitted information be correctly decoded
at the receiver with vanishing error probability, while in some real
systems it is acceptable to lose a small portion of the transmitted
information as long as there is a mechanism to cope with the packet
loss. For example, in systems with a receiver complexity constraint,
decoding over finite blocklength is necessary but in the case of
packet loss, ARQ (automatic repeat request) protocols are
implemented where the receiver requests retransmission of the lost
information \cite{Caire2001,Ghanim06}. If the system has a stringent
delay constraint, lost information can be approximated from the
context, for example the block-coded JPEG image transmission over
noisy channels where missing blocks can be reconstructed in the
frequency domain by interpolating the discrete cosine transformation
(DCT) coefficients of available neighboring blocks \cite{Ancis1999}.
These examples demonstrate a new notion of {\em capacity versus
outage}: the transmitter sends information at a fixed rate, which is
correctly received most of the time; with some maximal probability
$q$, the decoder sees a bad channel and declares an outage, and the
transmitted information is lost. This is formalized in the following
definition:
\begin{definition}[Capacity versus Outage]
\label{defn:capversusoutage} {\em Consider a composite channel
$\bs{W}$ with CSIR. A $(2^{nR},n)$ channel code for $\bs{W}$
consists of the following:
\begin{itemize}
\item an encoding function $X^n:
\mathcal{U} = \{1,2,\cdots,2^{nR}\}\to\mathcal{X}^n$, where
$\mathcal{U}$ is the message index set and $\mathcal{X}$ is the
input alphabet;
\item an outage identification function $I: \mathcal{S} \to \{0,1\}$,
where $\mathcal{S}$ is the set of channel states;
\item a decoding function $g_n:
\mathcal{Y}^n \times \mathcal{S} \to \hat{\mathcal{U}} = \{1, 2,
\cdots, 2^{nR}\}$, which only operates when $I=1$.
\end{itemize}
Define the outage probability
\[
P_o^{(n)} = \Pr\{I=0\}
\] and the error probability in non-outage states
\[
P_e^{(n)} = \Pr \{ U \neq \hat{U}| I=1 \}.
\]
A rate $R$ is outage-$q$ achievable if there exists a sequence of
$(2^{nR},n)$ channel codes such that ${\displaystyle
\lim_{n\rightarrow\infty}P_o^{(n)}\leq q }$ and ${\displaystyle
\lim_{n\rightarrow\infty}P_e^{(n)} = 0}$. The} capacity versus
outage $C_q$ {\em of the channel $\bs{W}$ with CSIR is defined to be
the supremum over all outage-$q$ achievable rates.}
\end{definition}

In the above definition, $P_o^{(n)}$ is the probability that the
decoder, using its side information about the channel, determines it
cannot reliably decode the received channel output and declares an
outage. In contrast, $P_e^{(n)}$ is the probability that the
receiver decodes improperly given that an outage is not declared.
Definition \ref{defn:capversusoutage} can be viewed as an {\em
operational definition} of the capacity versus outage. In parallel
to the development of the Shannon capacity, we also give an {\em
information theoretic definition} \cite[p. 184]{CoverIT} of the
capacity versus outage
\begin{eqnarray}
C_q  &=&  \sup_{\bs{X}} \ubar{\bs{I}}_q(\bs{X};\bs{Y}|S)
\nonumber \\
&=& \sup_{\bs{X}} \sup \left\{ \alpha:
F_{\bs{X}}(\alpha) \leq q \right\}. \label{eqn:outage}
\end{eqnarray}
Notice that $C_0=C$, so the capacity versus outage  is a
generalization of the Shannon capacity. The achievability proof
follows the same typical-set argument given in Section
\ref{sec:Shannon}. The converse result likewise follows
\cite{VerduH:94}. Details are given in Appendix \ref{app:outage}.

The concept of capacity versus outage was initially proposed in
\cite{OzarowS:94} for cellular mobile radios. See also \cite[Ch.
4]{Goldsmith} and references therein for more details. A
closely-related concept of $\epsilon$-capacity was defined in
\cite{VerduH:94}. However, there is a subtle difference between the
two: in the definition of $\epsilon$-capacity the non-zero error
probability $\epsilon$  accounts for decoding errors {\em
undetected} at the receiver. In contrast, in the definition of
capacity versus outage the receiver declares an outage when the
channel state does not allow the receiver to decode with vanishing
error probability. Asymptotically, the probability of error must be
bounded by some fixed constant $q$ and all errors must be {\em
recognized} at the decoder. As a consequence, no decoding is
performed for outage states. If the power consumption to perform
receiver decoding becomes an issue, as in the case of sensor
networks with non-rechargeable nodes or power-conserving mobile
devices, then we should distinguish between decoding with error and
no decoding at all in view of energy conservation.

This subtle difference also has important consequences when we
consider end-to-end communication performance using source and
channel coding. When the outage states are recognized by the
receiver, it can request a retransmission or simply reconstruct the
source symbol by its mean -- giving an expected distortion equal to
the source variance. In contrast, if the receiver cannot recognize
the decoding error as in the case of an $\epsilon$-capacity channel
code, the reconstruction based on the incorrectly decoded symbol may
lead to not only large distortion but also loss of synchronization
in the source code's decoder.

We can further define the {\em outage capacity} $C^o_q = (1-q)C_q$
as the long-term average rate, if the channel is used repeatedly and
at each use the channel state is drawn independently according to
$p(s)$. The transmitter uses a single codebook and sends information
at rate $C_q$; the receiver can correctly decode the information a
proportion $(1-q)$ of the time and turns itself off a proportion $q$
of the time. The outage capacity $C^o_q$ is a meaningful metric if
we are only interested in the fraction of correctly received packets
and approximate the unreliable packets by surrounding samples. In
this case, optimizing over the outage probability $q$ to maximize
$C^o_q$ guarantees performance that is at least as good as the
Shannon capacity and may be far better. As another example, if all
information must be correctly decoded eventually, the packets that
suffer an outage have to be retransmitted. This demands some
repetition mechanism that is usually implemented in the link-layer
error control of data communication. The number of channel uses $K$
to transmit a packet of size $(N=C_q)$ bits has a geometric
distribution
\[
\Pr \{K= k \} = q^{k-1}(1-q),
\]
and the expected value is $\frac{1}{(1-q)}=\frac{N}{C^o_q}$, which
also illustrates $C^o_q$ as a measure of the long-term average
throughput.

Next we briefly analyze the capacity versus outage from a
computational perspective. We need the following definition before
we proceed:
\begin{definition}[Probability-$q$ Compatible Subchannel]
{\em Consider a composite channel $\bs{W}$ with state distribution
$p(s)$, $s \in \mathcal{S}$. Consider another channel $\bs{W}_q$
where the channel state set ${\cal S}_q$ is a subset of ${\cal S}$
($\mathcal{S}_q \subseteq \mathcal{S}$). $\bs{W}_q$ is a}
probability-$q$ compatible subchannel {\em of $\bs{W}$ if
$\Pr\{\mathcal{S}_q\} \ge 1-q$.}
\end{definition}
Note that $\bs{W}_q$ is not exactly a composite channel since we
only specify the state set $\mathcal{S}_q$ but not the corresponding
state distribution over $\mathcal{S}_q$. However, we will only be
interested in the Shannon capacity of $\bs{W}_q$, and as pointed out
by Lemma \ref{lemma:sameShannonC}, the exact distribution over
$\mathcal{S}_q$ is irrelevant to determine this capacity.

The capacity versus outage as defined in \eqref{eqn:outage} requires
a two-stage optimization. In the first step we fix the input
distribution $\bs{X}$ and find the probability-$q$ compatible
subchannel that yields the highest achievable rate. In the second
step we optimize over the distribution of $\bs{X}$. This view is
more convenient if the optimal input distribution can be easily
determined. We then evaluate the achievable rate of each component
channel with this optimal input and declare outage for those with
the lowest rates. As an example, consider a slow-fading MIMO channel
with $m$ transmit antennas. Assume the channel matrix $\bs{H}$ has
i.i.d. Rayleigh fading coefficients. The outage probability
associated with transmit rate $R$ is known to be \cite{Telatar99}
\[
P_o(R) = \inf_{\bs{Q}\succeq0, \textrm{Tr}(\bs{Q})\le m} \Pr \left[
\log\det\ \left( \bs{I}+
\frac{\textsf{SNR}}{m}\bs{H}\bs{Q}\bs{H}^\dagger \right) \le R
\right],
\]
and the capacity versus outage is $C_q = \sup \{R: P_o(R) \le q\}$.
Although the optimal input covariance matrix $\bs{Q}$ is unknown in
general, it is shown in \cite{Zheng02} that there is no loss of
generality in assuming $\bs{Q}=\bs{I}$ in the high SNR regime and
the corresponding capacity versus outage simplifies to
\[
C_q = \sup \left\{R:  \Pr \left[ \log\det\ \left( \bs{I}+
\frac{\textsf{SNR}}{m}\bs{H}\bs{H}^\dagger \right) \le R \right] \le
q \right\}.
\]

By reversing the order of the two optimization steps we have another
interpretation of capacity versus outage
\begin{equation}
C_q = \sup_{\bs{W}_q
} C(\bs{W}_q).
\label{eqn:alterOutageCap}
\end{equation}
Here we first determine the Shannon capacity of each probability-$q$
compatible subchannel, then optimize by choosing the one with the
highest Shannon capacity. This view highlights the connection
between $C_q$ of a composite channel and the Shannon capacity of its
probability-$q$ compatible subchannels, and is more convenient if
there is an intrinsic ``ordering" of the component channels. For
example consider a degraded collection of channels where for any
channel states $s_1$ and $s_2$ there exists a transition probability
$p(y_2^n|y_1^n)$ such that
\[
p(y_2^n|x^n,s_2) = \sum_{y_1^n} p(y_1^n|x^n,s_1) p(y_2^n|y_1^n).
\]
The degraded relationship can be extended to the {\em less noisy}
and {\em more capable} conditions \cite{Elgamal_classofBC}. The more
capable condition requires\footnote{Assuming each component channel
is stationary and ergodic, the mutual information in
\eqref{eqn:morecapable} is well defined.}
\begin{equation}
I(X^n;Y_1^n|s_1) \ge I(X^n;Y_2^n|s_2) \label{eqn:morecapable}
\end{equation}
for any input distribution $\bs{X}$. It is the weakest of all three
but suffices to establish an ordering. The optimal probability-$q$
compatible subchannel $\bs{W}_q^*$ has the smallest set of channel
states ${\cal S}_q^*$ such that any component channel within ${\cal
S}_q^*$ is more capable than a component channel not in ${\cal
S}_q^*$. The Shannon capacity of $\bs{W}_q^*$ equals the capacity
versus outage-$q$ of the original channel $\bs{W}$.

\section{Expected Capacity} \label{sec:expectedCap}
The definition of capacity versus outage in \refS{outageCapacity} is
essentially an all-or-nothing game: the receiver may declare outage
for undesirable channel states but is otherwise required to decode
all transmitted information. There are examples where {\em partial}
received information is useful. Consider sending a multi-resolution
source code over a composite channel. Decoding all transmitted
information leads to reconstruction with the lowest distortion.
However, in the case of inferior channel quality, it still helps to
decode partial information and get a coarse reconstruction. Although
the transmitter sends information at a fixed rate, the notion of
expected capacity allows the receiver to decide in expectation how
much information can be correctly decoded based on channel
realizations.


Next we introduce some notation which is useful for the formal
definition of the expected capacity. Conventionally we represent
information as a message index, c.f. the Shannon capacity definition
\cite[p. 193]{CoverIT} and the capacity versus outage definition in
\refS{outageCapacity}. To deal with partial information, here we
represent information as a block of bits $(b_i)_{i \in
\mathcal{I}}$,
where $\mathcal{I}$ is
the set of bit indices. Denote by
\[
\mathcal{M}(\mathcal{I}) = \left\{ (b_i)_{i \in \mathcal{I}}: b_i
\textrm{ binary}\right\}
\]
the set of all possible blocks of information bits with bit indices
from the set $\mathcal{I}$. Each element in ${\cal M}({\cal I})$ is
a bit-vector of length $|{\cal I}|$, so the size of the set ${\cal
M}({\cal I})$ is $2^{|\mathcal{I}|}$. If another index set
$\widetilde{\cal I}$ is a proper subset of ${\cal I}$
($\widetilde{\cal I} \subset {\cal I}$), then ${\cal
M}(\widetilde{\cal I})$ represents some partial information with
respect to the full information ${\cal M}({\cal I})$. This
representation generalizes the conventional representation using
message indices.
\begin{definition}[Expected Capacity]
\label{defn:expectedC} Consider a composite channel $\bs{W}$ with
channel state distribution $p(s)$. {\em A $(2^{nR_t}, \{2^{nR_s}\},
n)$ code consists of the following:
\begin{itemize}
\item an encoding function
\[
f_n: \mathcal{M}(\mathcal{I}_{n,t}) = \{(b_i)_{i \in
\mathcal{I}_{n,t}} \} \to \mathcal{X}^n,
\]
where $\mathcal{I}_{n,t} = \{1,2,\cdots, nR_t\}$ is the index set of
the transmitted information bits and $\mathcal{X}$ is the input
alphabet;
\item a collection of decoders, one for each channel state $s$,
\[
g_{n,s}: \mathcal{Y}^n \times \mathcal{S} \to
\mathcal{M}(\mathcal{I}_{n,s}) = \{(\hat{b}_i)_{i \in
\mathcal{I}_{n,s}} \}
\]
where $\mathcal{I}_{n,s} \subseteq \mathcal{I}_{n,t}$ is the set of
indices of the decodable information bits in channel state $s$.
$|\mathcal{I}_{n,s}| = nR_s$.
\end{itemize}
Define the decoding error probability associated with channel state
$s$ as
\[
P_e^{(n,s)} = \Pr \left\{ \cup_{i \in \mathcal{I}_{n,s}} (\hat{b}_i
\neq b_i) \right\},
\]
and the average error probability
\[
P_e^{(n)} = \mathbb{E}_S P_e^{(n,S)} = \int P_e^{(n,s)} p(s)ds.
\]
A rate $R = \mathbb{E}_S R_S$ is achievable in expectation if there
exists a sequence of $(2^{nR_t}, \{2^{nR_s}\}, n)$ codes with
average error probability ${\displaystyle \lim_{n \to \infty}
P_e^{(n)} = 0}$. The} expected capacity $C^e(\bs{W})$ {\em is the
supremum of all rates $R$ achievable in expectation.}
\end{definition}
We want to emphasize a few subtle points in the above definition. In
channel state $s$ the receiver only decodes those information bits
$(b_i)$ with indices $i\in \mathcal{I}_{n,s}$. Decoding error occurs
if any of the decoded information bits $(\hat{b}_i)$ is different
from the transmitted information bit $(b_i)$. No attempt is made to
decode information bits with indices out of the index set
$\mathcal{I}_{n,s}$; hence these information bits are irrelevant to
the error analysis for channel state $s$.

The cardinality $nR_s$ of the index set $\mathcal{I}_{n,s}$ depends
only on the blocklength $n$ and the channel state $s$. Among the
transmitted $nR_t$ information bits, the transmitter and the
receiver can agree on the set of decodable information bits for each
channel state before transmission starts, i.e. not only the
cardinality of $\mathcal{I}_{n,s}$, but the set $\mathcal{I}_{n,s}$
itself is uniquely determined by the channel state $s$.
Nevertheless, for the same channel state $s$, the receiver may
choose to decode different sets of information bits depending on the
actual channel output $\mathcal{Y}^n$, although all these sets are
of the same cardinality $nR_s$. In this case the set of decodable
information bits for each channel state is unknown to the
transmitter beforehand.

We first look at the case where the transmitter and the receiver
agree on the set of decodable information bits for each channel
state. In a composite channel the transmitter can view the channel
as a broadcast channel with a collection of virtual receivers
indexed by channel realization $S$. The encoder uses a broadcast
code to transmit to the virtual receivers. The receiver uses the
side information $S$ to choose the appropriate decoder. Before we
proceed to establish a connection between the expected capacity of a
composite channel and the capacity region of a broadcast channel, we
state the following definition of the broadcast capacity region,
which is a direct extension from the two-user case \cite[p.
421]{CoverIT} to the multi-user case.

Consider a broadcast channel with $m$ receivers. The receivers are
indexed by the set $\mathcal{S}$ with cardinality $m$, which
is reminiscent of the index set of channel states in a composite
channel. The power set $\mathcal{P}(\mathcal{S})$ (or simply
$\mathcal{P}$) is the set of all subsets of $\mathcal{S}$. The
cardinality of the power set is $|\mathcal{P}(\mathcal{S})|=2^m$.

\begin{definition}[Broadcast Channel Capacity Region]
\label{defn:BCdefn} {\em A $(\{2^{nR_p}\},n)$ code for a broadcast
channel consists of the following:
\begin{itemize}
\item an encoder
\[
f_n: \,\, \prod_{ p \in \mathcal{P}, \,\,  p \neq \phi}
\mathcal{M}_p \to \mathcal{X}^n,
\]
where $\phi$ is the empty set, $p\in \mathcal{P}(\mathcal{S})$ is a
non-empty subset of users, and $\mathcal{M}_p = \{1,2,\cdots,
2^{nR_p} \}$ is the message set intended for users within the subset
$p$ only. The short-hand notation  $\prod_p \mathcal{M}_p$ denotes
the Cartesian product of the corresponding message sets;
\item a collection of $m$ decoders, one for each user $s$,
\[
g_{n,s}: \mathcal{Y}_s^n \to \prod_{p\in \mathcal{P}, \,\, s\in p}
\hat{\mathcal{M}}_p,
\]
where $\mathcal{Y}_s^n$ is the channel output for user $s$.
\end{itemize}
Define the error event $E_s$ for each user as
\begin{equation}
E_s = \left\{ g_{n,s}(Y_s^n) =  \left( \hat{M}_p \right)_{p\in
\mathcal{P}: \, s\in p}  \neq \big( M_p \big)_{p\in \mathcal{P}: \,
s\in p} \right\}, \label{eqn:Es}
\end{equation}
and the overall probability of error as
\[
P_e^{(n)} = \Pr \left\{ \bigcup_s E_s  \right\}.
\]
A rate vector $\{R_p\}_{p \in \mathcal{P}}$ is broadcast achievable
if there exists a sequence of $(\{2^{nR_p}\}, n)$ codes with
${\displaystyle \lim_{n \to \infty} P_e^{(n)} = 0}$. The} broadcast
channel capacity region $\mathcal{C}_{\BC}$ {\em is the convex
closure of all broadcast achievable rate vectors.}
\end{definition}
In the above definition, we explicitly distinguish between private
and common information. The message set $\mathcal{M}_p$ contains
information decodable by all users $s \in p$ but no others. For
instance, in a three-user BC we have private information
$\mathcal{M}_1$, $\mathcal{M}_2$, $\mathcal{M}_3$, information for
any pair of users $\mathcal{M}_{12}$, $\mathcal{M}_{23}$,
$\mathcal{M}_{13}$, and the common information $\mathcal{M}_{123}$.
The total number of message sets is $2^m-1$ since the empty set
$\phi$ is excluded.

We establish a connection between the expected capacity of a
composite channel and the capacity region of a broadcast channel
through the following theorem. For ease of notation we state the
theorem for a finite number of users (channel states). The result
can be generalized to an infinite number of users (continuous
channel state alphabets) using the standard technique of \cite[Ch.
7]{Gallagerbook}, i.e. to first discretize the continuous channel
state distribution and then take the limiting case.
\begin{theorem}
\label{thm:CeBCmapping} {\em Consider a composite channel
characterized by the joint distribution
\[
P_{W^n}(s, y^n|x^n) =  P_S(s) P_{Y^n|X^n,S}(y^n|x^n,s),
\]
and the corresponding BC
with the channel for each receiver satisfying
\[
P_{Y_s^n|X^n}(y_s^n|x^n) = P_{Y^n|X^n,S}(y_s^n|x^n,s).
\]
Denote by $C^e$ the expected capacity of the composite channel and
by $\mathcal{C}_{\BC}$ the capacity region of the corresponding BC,
as in Definitions \ref{defn:expectedC} and \ref{defn:BCdefn},
respectively. If the set of decodable information bits in the
composite channel is uniquely determined by the channel state $S$,
then the expected capacity satisfies}
\begin{equation} C^e = \sup_{(R_p) \in \mathcal{C}_{\BC}}
\sum_{p \in \mathcal{P} } R_p \sum_{s\in p} P_S(s) = \sup_{(R_p) \in
\mathcal{C}_{\BC}} \sum_{ s \in \mathcal{S} } P_S(s) \sum_{s\in p}
R_p. \label{eqn:CeBCTheorem}
\end{equation}
\end{theorem}
The proof establishes a two-way mapping: any $(\{2^{nR_p}\}, n)$
code for the broadcast channel can be mapped to a $(2^{nR_t},
\{2^{nR_s}\}, n)$ expected-rate code for the composite channel and
vice versa, where the mapping satisfies $R_s = \sum_{s\in p}R_p$ for
channel state $s$. The details are given in Appendix
\ref{app:mapping}.

Although we have introduced a new notion of capacity, the connection
established in Theorem \ref{thm:CeBCmapping} shows that the tools
developed for broadcast codes can be applied to derive corresponding
expected capacity results, with the addition of an optimization to
choose the point on the BC rate region boundary that maximizes the
expected rate. For example, in \cite{Shamai03} some suboptimal
approaches, including super-majorization and one-dimensional
approximation, were introduced to analyze the expected capacity of a
single-user slowly fading MIMO channel. After the full
characterization of the MIMO BC capacity region through the work
\cite{Caire2003,SriramDuality2003,WeiYuBC2004,Viswanath03,Weingarten06},
the expected capacity of a slowly fading MIMO channel can be
obtained by choosing the optimal operating point on the boundary of
the dirty-paper coding (DPC) region.

The connection in Theorem \ref{thm:CeBCmapping} also shows that any
expected-rate code designed for a composite channel can be put into
the framework of BC code design. Strategies like {\em layered source
coding with progressive transmission}, proposed in \cite{Gunduz05},
immediately generalize to the broadcast coding problem. Assuming
there are only two channel states $s_1$ and $s_2$, this strategy
divides the entire transmission block into two segments. The
information transmitted in the first segment is intended for both
states, and that in the second segment is intended for the better
channel state $s_2$ only. This strategy can be easily mapped to a BC
code with individual information $\mathcal{M}_2$ and common
information $\mathcal{M}_{12}$, and orthogonal channel access.
Furthermore, the complexity of deriving a single point on the BC
region boundary is similar to that of deriving the expected capacity
under a specific channel state distribution. The entire BC region
boundary can be traced out by varying the channel state
distributions.

We want to emphasize that in Theorem \ref{thm:CeBCmapping} the
condition that the transmitter knows the set of decodable
information bits in advance is not superfluous. If the receiver
chooses to decode different sets of information bits depending on
the actual channel output $\mathcal{Y}^n$, and consequently the
transmitter does not know the set of decodable information bits for
each state $s$, then the mapping between expected-rate codes and BC
codes may not exist. In the following we give an example where the
expected capacity exceeds the supremum of expected rates achievable
by BC codes. Consider a binary erasure channel (BEC) where the
erasure probability takes two equiprobable values $0 \le \alpha_1 <
\alpha_2 \le 1$. In Appendix \ref{app:BECBC} we show that the
maximum expected rate achievable by BC codes is
\begin{equation}
R = \max\left\{ 1-\alpha_2, \frac{1-\alpha_1}{2} \right\}.
\label{eqn:BECBCR}
\end{equation}
However, we can transmit uncoded information bits directly over this
composite BEC. In the limit of large blocklength $n$, the receiver
can successfully decode $n(1-\alpha_i)$ bits for channel states
$\alpha_i$, $i=1,2$, by simply inspecting the channel output,
although these successfully decoded information bits cannot be
determined at the transmitter {\em a priori}. Overall the expected
capacity
\[
C^e = 1 - \frac{\alpha_1 + \alpha_2}{2}
\]
exceeds the maximum expected rate achievable by BC codes. Notice,
however, these two channel codes are extremely different from an
end-to-end coding perspective. The broadcast strategy may be
combined with a multiresolution source code. In contrast, the source
coding strategy required for the uncoded case is a multiple
description source code with single-bit descriptions. Due to this
difference, it is not obvious which scenario yields the lower
end-to-end distortion. The comparison depends on the channel state
distribution and the rate-distortion function of the source.

Regardless of the transmitter's knowledge about decodable
information bits, we show that $C^e$ satisfies the lower bound $C^e
\ge \sup_q C^o_q$ and the upper bound
\begin{equation}
C^e \le \sup_{\bs{X}}\limsup_{n\rightarrow\infty}\mathbb{E}_S
\mathbb{E}_{X^nY^n|S} \left[ \left. \frac{1}{n}i_{X^nW^n}(X^n;Y^n|S)
\right|S \right]. \label{eqn:ce}
\end{equation}
The lower bound is achieved using the channel code for capacity
versus outage-$q$, which achieves a rate $C_q$ a proportion $(1-q)$
of the time and zero otherwise. For the upper bound, we assume
channel side information is provided to the transmitter (CSIT) so it
can adapt the transmission rate to the channel state. In this case,
the achievable expected rate can only be improved. The proof is
given in Appendix \ref{app:upperbound}.

\section{Examples}\label{sec:ex}
In this section we consider some examples to illustrate various
capacity definitions.
\subsection{Gilbert-Elliott Channel} The Gilbert-Elliott channel
\cite{MushkinB:89} is a two-state Markov chain, where each state is
a BSC as shown in \refF{GEchannel}. The crossover probabilities for
the ``good'' and ``bad'' BSCs satisfy $0\le p_G<p_B \le {1/2}$. The
transition probabilities between the states are $g$ and $b$
respectively. The initial state distribution is given by $\pi_G$ and
$\pi_B$ for states $G$ and $B$.  We let
 $x_{n}\in\{0,1\}$, $y_{n}\in\{0,1\}$, and
$z_{n}=x_{n}\oplus y_{n}$ denote the channel input, output, and
error on the $n$th transmission. We then study capacity definitions
when the channel characteristics of stationarity and ergodicity
change with the parameters.\\ \Fig{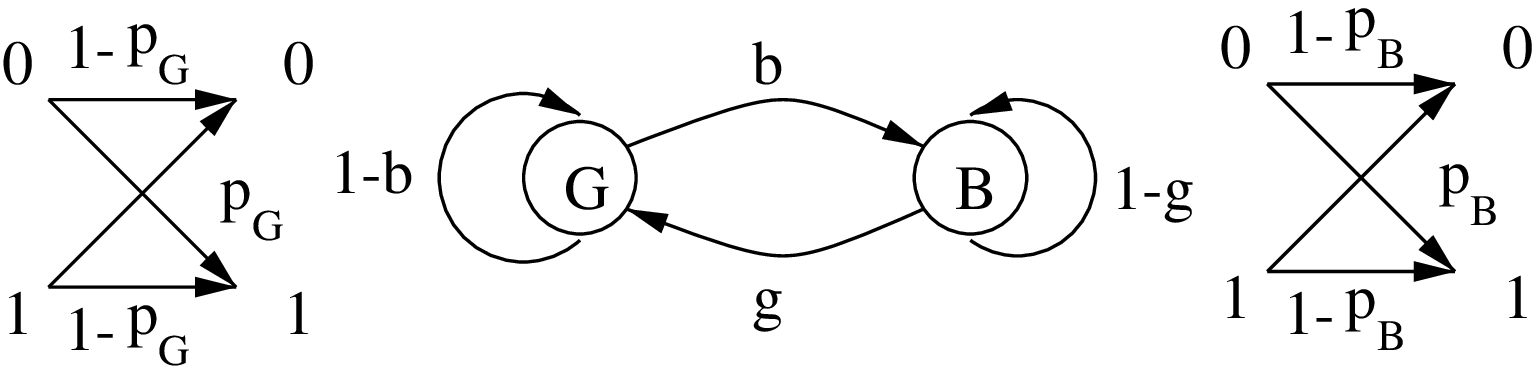}{Gilbert-Elliott
Channel}

\noindent {\em Example 1: Ergodic Case, Stationary or Non-Stationary}\\
When $\pi_G=g/(g+b)$ and $\pi_B=b/(g+b)$, the Gilbert-Elliott
channel is stationary and ergodic. In this case the information
density $\frac{1}{n}i_{X^nW^n}(X^n;Y^n)$ converges to a
$\delta$-function at the average mutual information, so capacity
equals average mutual information as usual. Therefore the Shannon
capacity $C$ is equal to the expected capacity $\pi_G C_G + \pi_B
C_B$, where $C_G=1-h(p_G)$, $C_B=1-h(p_B)$ and $h(p) = -p \log p
-(1-p) \log (1-p)$ is the binary entropy function.

This is a single-state composite channel. Since any transmission may
experience either a good or a bad channel condition, the receiver
has no basis for choosing to declare an outage on certain
transmissions and not on others. Capacity versus outage equals
Shannon capacity in this case.

If $\pi_G \neq g/(g+b)$ but $b$ and $g$ are nonzero, then the
Gilbert-Elliott channel is ergodic but not stationary. However, the
distribution on the states $G$ and $B$ converges to a stationary
distribution. Thus the channel is asymptotically mean stationary,
and the definitions of capacity have the same values as in the
stationary case. \\

\noindent{\em Example 2: Stationary and Nonergodic Case}\\
We now set $g=b=0$. So the initial channel state is chosen according
to probabilities $\{\pi_G,\pi_B\}$ and then remains fixed for all
time. The Shannon capacity equals that of the bad channel $(C=C_B)$.
The capacity versus outage-$q$ $C_q = C_B$ if the outage probability
$q<\pi_B$ and $C_q = C_G$ otherwise.
%
The loss incurred from lack of side information at the encoder is
that the expected capacity is strictly less than the average of
individual capacities $\pi_B C_B + \pi_G C_G$ and is equal to
\cite{Cover72}
\begin{equation}
\max_{0\le r \le 1/2} 1-h(r*p_B) + \pi_G [ h(r*p_G) - h(p_G) ],
\label{eqn:BSC2}
\end{equation}
where $\alpha* \beta = \alpha(1-\beta) + (1-\alpha) \beta$.  The
interpretation here is that the broadcast code achieves rate
$1-h(r*p_B)$ for the bad channel and an additional rate $h(r*p_G) -
h(p_G)$ for the good channel, so the average rate is the expected
capacity.

Using the Lagrangian multiplier method we can obtain $r^*$ which
maximizes \eqref{eqn:BSC2}. Namely if we define
\[
k = \frac{\pi_G}{\pi_B}, \quad  A = \frac{1-2p_B}{1-2p_G}, \quad
f(p_1,p_2) = \frac{ \log(1/p_1-1)} {\log(1/p_2-1)}
\]
then $r^*=0$ if $k \le Af(p_B,p_G)$; $r^*=1/2$ if $k \ge A^2$ and
$r^*$ solves $f(r*p_G, r*p_B) = A/k$ otherwise.

\subsection{BSC with random crossover probabilities}
In the non-ergodic case, the Gilbert-Elliott Channel is a two-state
channel, where each state corresponds to a BSC with a different
crossover probability. We now generalize that example to allow more
than two states. We consider a BSC with random crossover probability
$0\le p \le 1/2$. At the beginning of time, $p$ is chosen according
to some distribution $f(p)$ and then held fixed. We also use $F(p) =
\int_0^p f(s)ds$ to denote the cumulative distribution function.
Like the non-ergodic Gilbert-Elliott channel, this is a multi-state
composite channel provided $\{p:f(p)>0\}$ has cardinality at least
two. The Shannon capacity is
$C = 1 - h(p^*) $
where \[ p^* = \sup \{ p: f(p) > 0 \} = \inf \{ p: F(p) =1 \},
\]
and the capacity versus outage-$q$ is
$C_q = 1 - h(p_q) $
where $p_q = \inf \{ p: F(p) \ge 1-q \}$.

We consider a broadcast approach on this channel to achieve the
expected capacity.
The receiver is equivalent to a continuum of ordered users, each
indexed by the BSC crossover probability $p$ and occurring with
probability $f(p)dp$. If the set $\{p:f(p)>0\}$ is infinite, then
the transmitter sends an infinite number of layers of coded
information and each user decodes an incremental rate $|dR(p)|$
corresponding to its own layer. Since the BSC broadcast channel is
degraded, a user with crossover probability $p$ can also decode
layers indexed by larger crossover probabilities, therefore we
achieve a rate of
\begin{equation}
R(p) = - \int_p^{1/2} dR(p)
\end{equation}
for receiver $p$. The problem of determining the expected capacity
then boils down to the characterization of the broadcast rate region
and the choice of the point on that region that maximizes $\int_p
R(p)f(p) dp$.

In the discrete case with $N$ users, assuming $0\le p_1 \le \cdots
\le p_N \le (1/2)$, the capacity region is shown to be
\cite{Bergmans73}
\begin{equation}
\left\{ \bs{R} = (R_i)_{1 \le i \le N}:  R_i = R(p_i) = h(r_i*p_i) -
h(r_{i-1}*p_i) \right\}
\end{equation}
where $0=r_0\le r_1 \le \cdots \le r_N = 1/2$. Since the original
broadcast channel is stochastically degraded it has the same
capacity region as a cascade of $N$ BSC's. The capacity region
boundary is traced out by augmenting $(N-1)$ auxiliary channels
\cite{Bergmans73} and varying the crossover probabilities of each.
For each $i$, $r_i$ equals the overall crossover probability for
auxiliary channels $1$ up to $i$. See \refF{BSC} for an
illustration. The resulting expected capacity is
\[
C^e = \max_{0=r_0\le \cdots \le r_N = 1/2} \sum_{i=1}^N f(p_i)
\sum_{j=i}^N [h(r_i*p_i) - h(r_{i-1}*p_i)]. \]
 \Fig{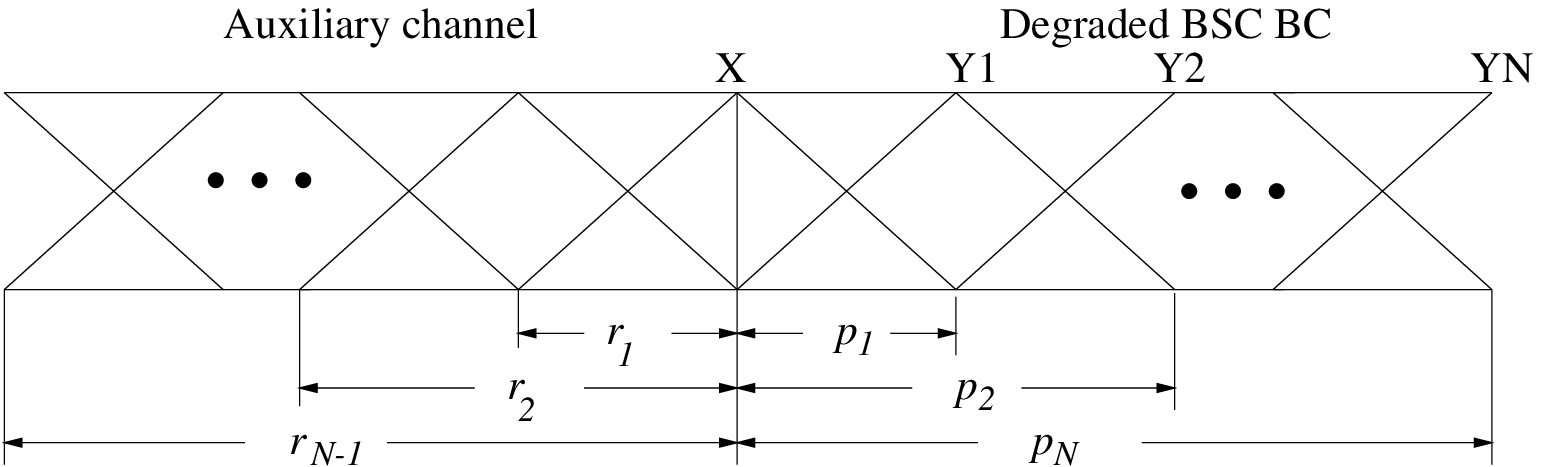}{BSC
broadcast channel with auxiliary channels for random coding}

We extend the above result to the continuous case with an infinite
number of auxiliary channels. In this case we define a monotonically
increasing function $r(p)$ equal to the overall crossover
probability of auxiliary channels up to that indexed by $p$. In the
following we use $r(p)$ and $r_p$ interchangeably. For the layer
indexed by $p$, the incremental rate is
\[
-dR(p) = h(p*r_p) - h(p*r_{p-dp}).
\]
Using the first order approximation $r_{p-dp} \approx r_p - r'_p dp$
and $h(x-\delta) \approx h(x) - h'(x)\delta$ for small $\delta$, we
obtain
\begin{eqnarray*}
-dR(p) &=& h(p*r_p) - h(p*r_{p-dp}) \\
&\approx& h(p*r_p) - h(p*r_p - (1-2p)r_p'dp) \\
&\approx& \log\left(\frac{1}{p*r_p}-1\right) (1-2p)r'_pdp,
\end{eqnarray*}
%
Note here $\delta = (1-2p)r_p'dp$ is a small variation, and we do
not explicitly address the problematic limiting case $h'(x)\to
\infty$ as $x$ approaches zero\footnote{The achievable rate $R(p)$
for any state is bounded by one, therefore
$\int_\epsilon^{1/2}f(p)R(p)dp$, as a function of $\epsilon$, is
right continuous at $\epsilon=0$. We can avoid the problematic
limiting case by focusing on strictly positive $\epsilon$ and obtain
the expected capacity \eqref{eqn:BSCexpectedCapacity} by
continuity.}.

Overall the expected rate is
\begin{eqnarray}
C^e &=& \int_0^{1/2} f(p)R(p)dp
= - \int_0^{1/2} F(p)dR(p) \nonumber \\
&=& \int_0^{1/2} F(p) \log\left(\frac{1}{p*r_p}-1\right)
(1-2p)r'_pdp. \label{eqn:BSCexpectedCapacity}
\end{eqnarray}
The optimal $r(p)$ maximizing the expected rate can be solved
through calculus of functional variation. Define $S(p,r_p,r'_p)$ as
\begin{equation}
S(p,r_p,r'_p) = F(p) \log\left(\frac{1}{p*r_p}-1\right) (1-2p)r'_p.
\end{equation}
The optimal $r(p)$ should satisfy the E\"{u}ler equation
\cite{Luenberger}
\begin{equation}
S_r - \frac{d}{dp}S_{r'} = 0 \label{eqn:Euler}
\end{equation}
where
\begin{eqnarray*}
&& S_r = \frac{\partial S}{\partial r} = -\frac{ (1-2p)^2 F(p) r'_p
}
{p*r_p-(p*r_p)^2}, \\
&& S_{r'} = \frac{\partial S}{\partial r'} = (1-2p) F(p)
\log\left[\frac{1-p*r_p}{p*r_p}\right], \\
&& \frac{dS_{r'}}{dp} = \left[ (1-2p) f(p) - 2F(p) \right]
\log\left[
\frac{1-p*r_p}{p*r_p} \right] \\
&& \quad - \frac{(1-2p) F(p)} {p*r_p - (p*r_p)^2} \left[ 1-2r_p +
(1-2p) r_p' \right].
\end{eqnarray*}
After some algebra \eqref{eqn:Euler} simplifies to
\begin{equation}
\frac{(p*r_p)^{-1} - (1-p*r_p)^{-1}}{\log(1-p*r_p)-\log(p*r_p)} =
\frac{(1-2p) f(p) - 2F(p)} {F(p)}. \label{eqn:soln}
\end{equation}
In general \eqref{eqn:soln} has no closed-form solution but there
exist obvious numerical approaches.

As an example, suppose that the crossover probability is uniformly
distributed on $[0, 1/2]$. The Shannon capacity is limited by the
worst channel state $(p=1/2)$, giving $C=0$. The capacity versus
outage-$q$ is $C_q = \left[1 - h(\frac{1-q}{2})\right]$. To
approximate the expected capacity, we solve for $r(p)$ in
\eqref{eqn:soln} for each $p$. It is seen that $0\le r_p \le 1/2$
only for $p_l \le p \le p_u$, where the two cutoff probabilities
satisfy $r(p_l) = 0$ and $r(p_u)=1/2$. For the uniform distribution
case, $p_l = 0.136$ and $p_u = 1/6$, which demonstrates that it is
unnecessary to use the channel all the time to achieve the expected
capacity. In fact no information is sent for $p \ge 1/6$.
\Fig{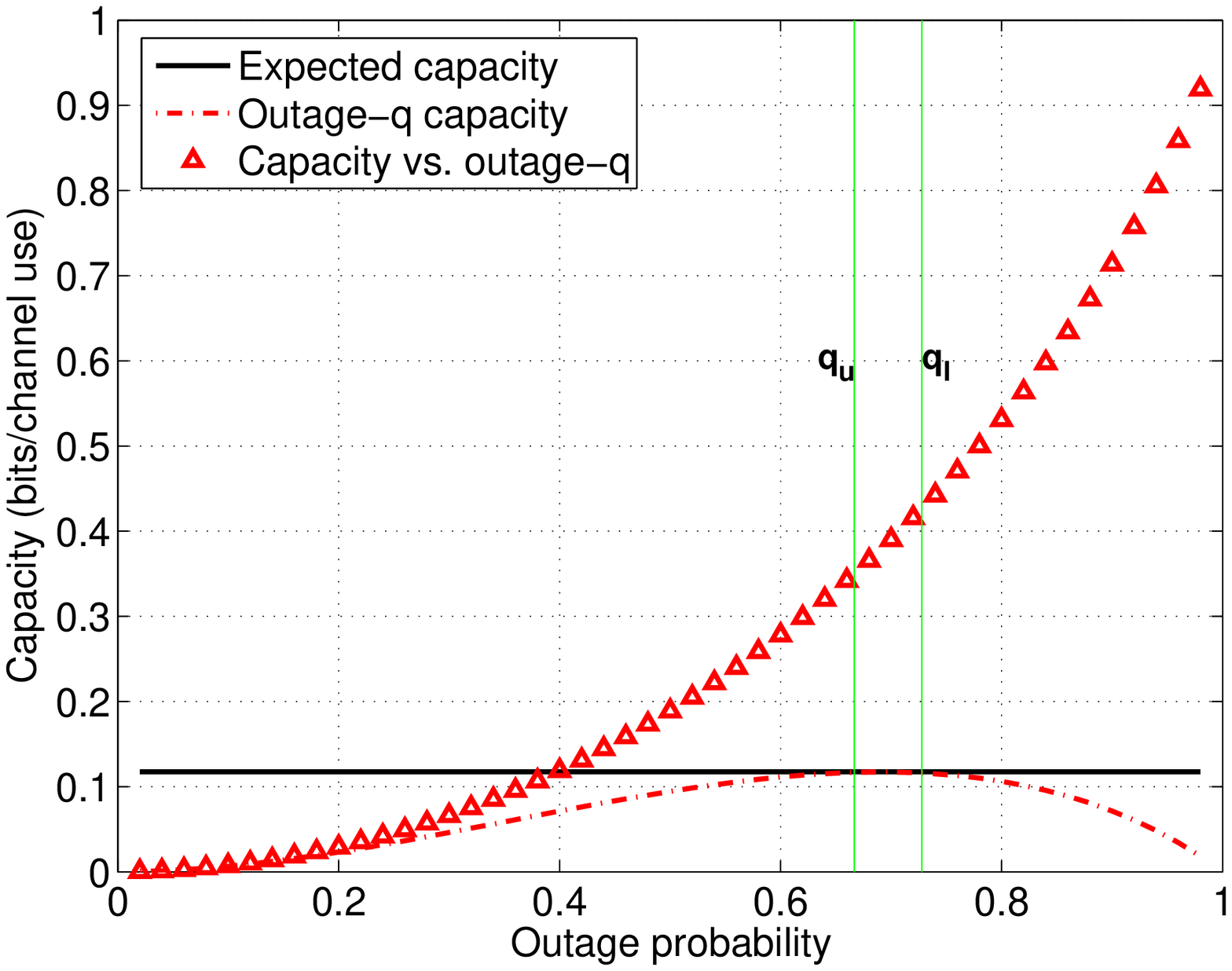}{Capacity under different definitions of BSC with random
crossover probability.} \Fig{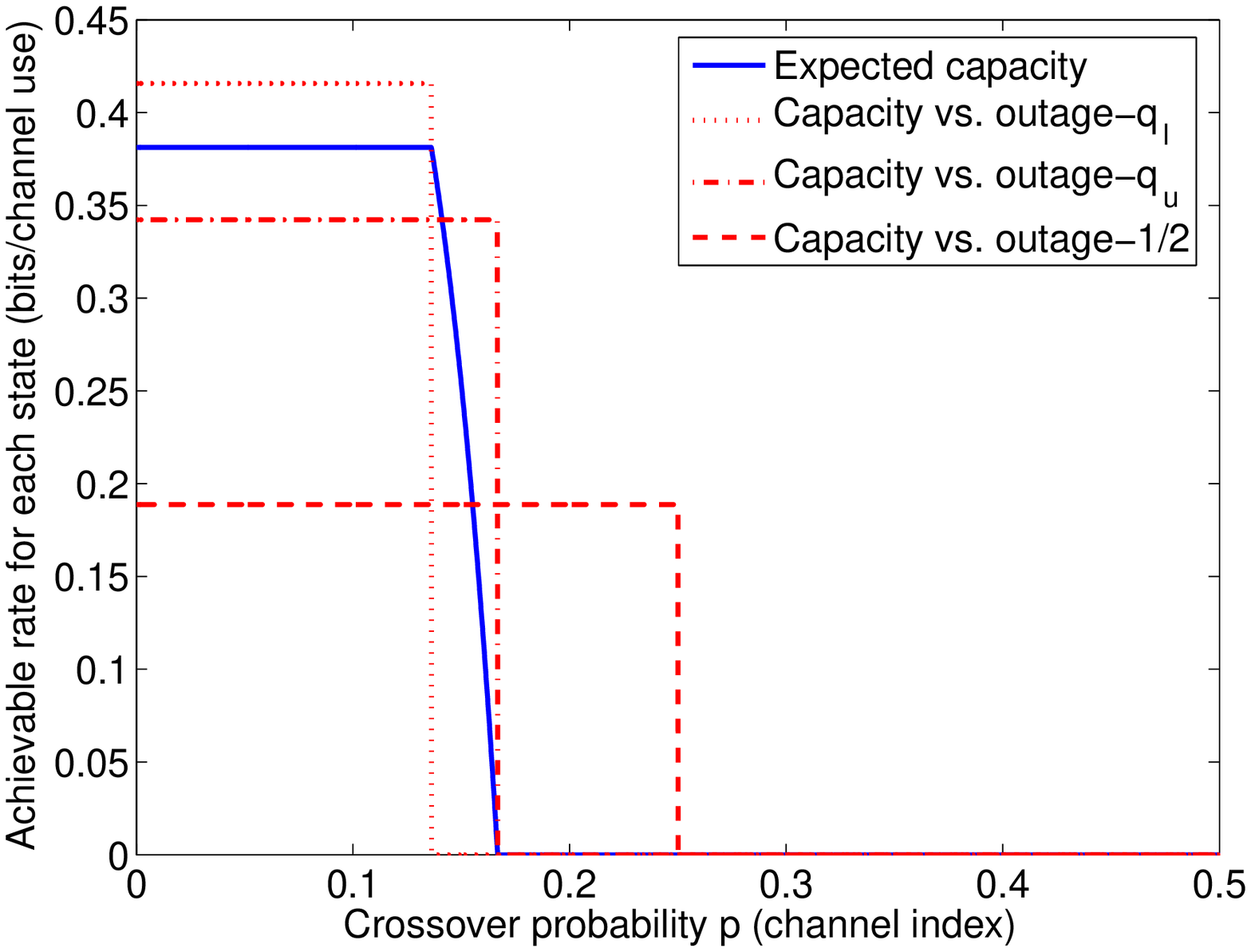}{Achievable rate for each
channel state} \Fig{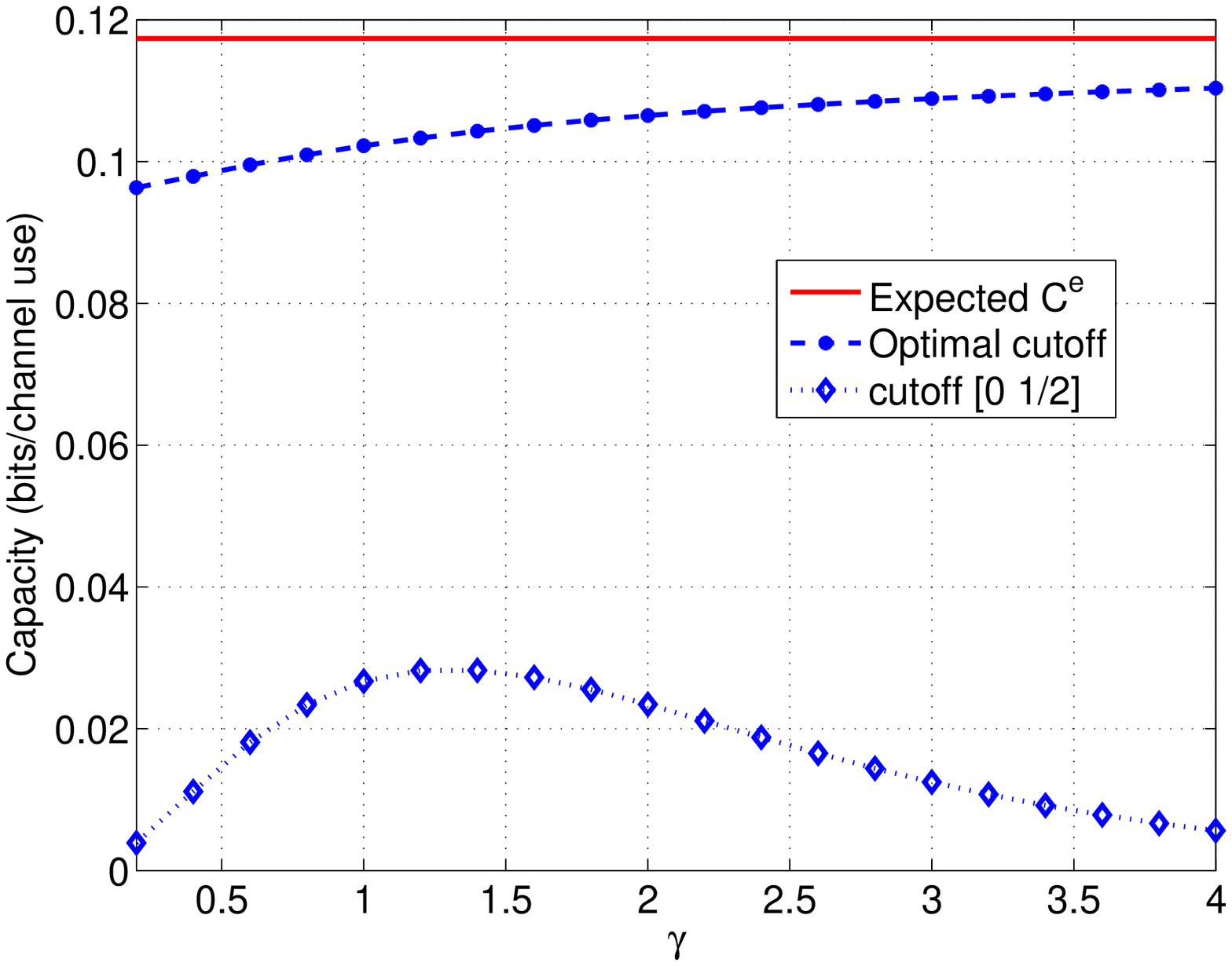}{Effect of cutoff range}

In \refF{cap} we plot the expected capacity, the outage-$q$
capacity, and the capacity versus outage-$q$. Although the capacity
versus outage-$q$ exceeds the expected capacity $C^e$ for some
values of $q$, the outage-$q$ capacity $C^o_q$ is always dominated
by the expected capacity $C^e$, since an outage-$q$ code is one of
many possible codes for the expected capacity. Define cutoff outage
probabilities $q_l = 1- 2p_l$ and $q_u = 1-2p_u$. Note that $C^o_q
\approx C^e$ for all $q \in [q_l, q_u]$. In this range an outage
code gives almost the same expected rate as a broadcast code.

In \refF{rateprofile} we plot the rate used in each state by the
expected capacity code and the capacity versus outage codes at
outage probabilities $q_l$, $q_u$ and $1/2$. We see that the code
for outage capacity achieves a constant rate for non-outage states
and a rate $0$ otherwise. For this example, the incremental rates
$|dR(p)|$ are nonzero only for $p_l \le p \le p_u$. Therefore the
code for expected capacity achieves a rate $0$ when $p>p_u$. As $p$
decreases from $p_u$ to $p_l$, the rate gradually increases from $0$
to $0.38$ bits per channel use, and stays at this constant level for
$p<p_l$. Since all channels are equally probable, the area under
each curve is the expected rate of that strategy. The area under the
expected capacity curve is the largest. The expected capacity curve
is, in some places, lower than the curve for outage-$q_l$ capacity.
Although the outage-$q_l$ code achieves a rate higher than the
broadcast code for expected capacity when $p<p_l$, the same code has
decoding rate $0$ for all other channel states $p>p_l$, giving a
lower area under the total curve.

A potential advantage of the outage code is its simplicity. The
transmission rate is fixed, so the code may be coupled with a
conventional source code. The advantage of the expected capacity
code is its higher expected rate. The code may be coupled with a
multiresolution source code. It is not obvious which strategy yields
better end-to-end coding performance in this example. In general, an
expected rate code is required to achieve the optimal end-to-end
distortion, but this code may use a rate vector on the boundary of
the BC capacity region which is different from the rate vector used
by the code that achieves the expected capacity \cite{Ng07b}.

The procedure to solve for the expected capacity is computationally
intensive. In the above example, when looking for the optimal $r(p)$
which leads to the expected capacity, we first identify the cutoff
probabilities $(p_l,p_u)$ and then solve \eqref{eqn:soln} for each
$p$ in this range. We want to emphasize that the correct cutoff
range, although seemingly a very coarse characterization of the
optimal solution, is crucial to the expected rate. Consider some
alternative approaches:
\begin{itemize}
\item Optimal cutoff $[p_l, p_u]$ with suboptimal $r(p)$:
\begin{equation}
r(p) = \left\{
\begin{array}{ll}
\frac{(p-p_l)^\gamma}{2(p_u-p_l)^\gamma}, & p_l \le p \le p_u, \\
0, & \textrm{otherwise}.
\end{array}
\right. \label{eqn:rp_opt}
\end{equation}

\item Cutoff range $[0,1/2]$:
\begin{equation}
r(p)= (1/2)(2p)^\gamma. \label{eqn:cutoff012}
\end{equation}
\end{itemize}
The choice of $\gamma$ makes $r(p)$ convex ($\gamma
>1$), linear ($\gamma = 1$) or concave ($\gamma <1$) in both
approaches. In \refF{random}, for $\gamma$ ranges between $0$ and
$4$ we plot the achievable expected rate using the cutoff range $[0,
1/2]$ and suboptimal $r(p)$ as in \eqref{eqn:cutoff012}, the
achievable expected rate using the optimal cutoff range $[p_l, p_u]$
and suboptimal $r(p)$ as in \eqref{eqn:rp_opt}, and the expected
capacity of this channel. We observe that the optimal cutoff range
yields an expected rate very close to $C^e$, but the expected rate
is clearly suboptimal if we use the cutoff range $[0,1/2]$. By
optimizing the cutoff range we actually capture most benefit of the
expected-rate code as compared to the conventional code for Shannon
capacity.

\section{Source-Channel Coding and Separation}
\label{sec:sccoding} Channel capacity theorems deal with data
transmission in a communication system. When extending the system to
include the source of the data, we also need to consider the data
compression problem.
For the overall system, the end-to-end distortion is a well-accepted
performance metric. When both the source and channel are stationary
and ergodic, codes are usually designed to achieve the same
end-to-end distortion level for any source sequence and channel
realization. However, if the channel model is generalized to such
scenarios as the composite channel above, it is natural to introduce
generalized end-to-end distortion metrics such as the {\em
distortion versus outage} and the {\em expected distortion}
\cite{Liang07ITW}, similar to the development of alternative
capacity definitions.
%
These alternative distortion metrics are also considered in prior
works \cite{Shamai98, Reznic06, Mittal02, Gunduz, Ng07b, Honig04}.

The renowned source-channel separation theorem \cite[Theorem
2.4]{CsiszarK:81} asserts that a target distortion level $D$ is
achievable if and only if the channel capacity $C$ exceeds the
source rate distortion function $R(D)$, and a two-stage separate
source-channel code suffices to meet the requirement\footnote{The
separation theorem for lossless transmission \cite{Shannon48} can be
regarded as a special case of zero distortion.}. This theorem
enables separate design of source and channel codes and guarantees
the optimal performance. However, there are a few underlying
assumptions: 
a single-user channel;
a stationary ergodic source and channel;
a single distortion level maintained for all transmission.
It is known that the separation theorem fails if the first two
assumptions do not hold \cite{ElGamal80, Vembu95}. In fact, the
end-to-end distortion metrics also dictate whether the
source-channel separation holds for a communication system. In
\cite{Liang07ITW} we showed the direct part of source-channel
separation under the distortion versus outage metric and established
the converse for certain systems. On the contrary, source-channel
separation does not hold under the expected distortion metric.

Source-channel separation implies that the operation of source and
channel coding does not depend on the statistics of the counterpart.
Meanwhile, the source and channel do need to communicate with each
other through an {\em interface}, which is a single number in the
classical separation theorem. For generalized source/channel models
and distortion metrics, the interface is not necessarily a single
rate and may allow multiple parameters to be agreed on between the
source and channel encoders and decoders. As we expect a performance
enhancement when source and channel exchange more information
through more sophisticated interface, an interesting topic for
future research would be to characterize the tradeoff between
interface complexity and the achievable end-to-end performance
\cite{LiangJournal2}.

\section{Conclusions}
\label{sec:con} In view of the pessimistic nature of Shannon
capacity for composite channels with CSIR, we propose alternative
capacity definitions including capacity versus and expected
capacity. These definitions lend insight to applications where side
information at the receiver combined with appropriate source coding
strategies can exploit these more flexible notions of capacity. We
prove capacity theorems or bounds under each definition, and
illustrate how expected achievable rates can be improved through
examples of Gilbert-Elliot channels and a BSC with random crossover
probabilities. While the use of capacity definitions inherently
focuses our attention on achievable (expected) rates, we note that
the existence of other meaningful measures of performance in the
given coding environment. For example, since outage-$q$ codes are
compatible with conventional source codes while expected capacity
codes require multiresolution or multiple description codes,
depending on whether or not the corresponding broadcast channel is
degraded, the fact that the expected rate of the expected capacity
code exceeds that of the outage-$q$ code does not guarantee lower
end-to-end expected distortion. Furthermore, since a non-ergodic
channel experiences a single ergodic mode for all time, there is
some justification for performance measures that take the
probability of suffering a very low-rate state into account. These
topics provide a wealth of interesting questions for future research
with some initial work presented in \cite{Ng07b, Gunduz,
Liang07ITW}.

\appendices
\section{Proof of Lemma \ref{lemma:sameShannonC}}
\label{app:ShannonC} We prove $C(\bs{W}_1) \le C(\bs{W}_2)$ if $p_1
\ll p_2$, and vice versa. Therefore equivalent probability measures
of $p_1$ and $p_2$ imply identical Shannon capacity. The result is
intuitive but we need to address a subtle technical issue: note that
$p_1$ and $p_2$ are channel state distributions, while the Shannon
capacity is defined through the information density distribution
\eqref{eqn:cdfFalpha}, which depends on both input and channel
statistics.

Recall the Shannon capacity formula \eqref{eqn:ShannonCapFormula}
\[
C(\bs{W}_1) =  \sup_{\bs{X}} \sup \{\alpha: F_{\bs{X}}(\alpha)=0\}.
\]
Denote by $\bs{X}_*$ the input distribution that achieves the
supremum in \eqref{eqn:ShannonCapFormula}, and by $F_1(\alpha)$ the
corresponding information density distribution.
For arbitrary $\epsilon > 0$, we define
\[
M_\epsilon(\alpha) = \left\{s: \lim_{n \to \infty}
 P_{X_*^nY^n|S} \left[ \frac1n
i_{X_*^nY^n|S}(X^n;Y^n|s) \le \alpha \right] \ge \epsilon \right\}.
\]
Notice that
\begin{eqnarray}
&& F_1(\alpha) \nonumber \\
&=& \lim_{n \to \infty} P_{X_*^nW_1^n} \left\{ \frac1n
i_{X_*^nW_1^n}(X^n;Y^n|S) \le
\alpha \right\} \nonumber \\
&=& \lim_{n \to \infty} \int  P_{X_*^nY^n|S} \left\{ \frac1n
i_{X_*^nY^n|S}(X^n;Y^n|s) \le \alpha \right\} \cdot p_1(s)ds \nonumber \\
&=& \int \lim_{n \to \infty} P_{X_*^nY^n|S} \left\{ \frac1n
i_{X_*^nY^n|S}(X^n;Y^n|s) \le \alpha \right\} \cdot p_1(s)ds \nonumber \\
&\ge& \epsilon \int_{M_\epsilon(\alpha)} p_1(s)ds,
\label{eqn:Mepsilonlowerbound}
\end{eqnarray}
where we exchange the order of integral and limit according to
dominant convergence theorem. From 
\eqref{eqn:Mepsilonlowerbound} we see that $F_1(\alpha)=0$ implies
\[
\int_{M_\epsilon(\alpha)} p_1(s)ds = 0.
\]
Assuming $p_1 \ll p_2$, it follows that
\[
\int_{M_\epsilon(\alpha)} p_2(s)ds = 0.
\]
Now define $F_2(\alpha)$ as the information density distribution of
channel $\bs{W}_2$ when evaluated at input $\bs{X}_*$, i.e.
\begin{eqnarray*}
&& F_2(\alpha) \\
&=& \lim_{n \to \infty} P_{X_*^nW_2^n} \left\{ \frac1n
i_{X_*^nW_2^n}(X^n;Y^n|S) \le \alpha \right\} \\
&=&  \int_{\mathcal{S}-M_\epsilon(\alpha)} \lim_{n \to \infty}
P_{X_*^nY^n|S} \left\{ \frac1n
i_{X_*^nY^n|S}(X^n;Y^n|s) \le \alpha \right\} \cdot p_2(s)ds \\
&& + \int_{M_\epsilon(\alpha)} \lim_{n \to \infty} P_{X_*^nY^n|S}
\left\{ \frac1n
i_{X_*^nY^n|S}(X^n;Y^n|s) \le \alpha \right\} \cdot p_2(s)ds \\
&\le& \epsilon \int_{\mathcal{S}-M_\epsilon(\alpha)} p_2(s)ds + \int_{M_\epsilon(\alpha)} p_2(s)ds \\
&\le& \epsilon.
\end{eqnarray*}
Since $\epsilon$ is arbitrary, we see that $F_1(\alpha)=0$ implies
$F_2(\alpha)=0$, therefore
\begin{eqnarray*}
C(\bs{W}_1) 
&=& \sup \{\alpha: F_1(\alpha)=0\} \\
&\le& \sup \{\alpha: F_2(\alpha)=0\} \\
&\le& C(\bs{W}_2).
\end{eqnarray*}

\section{Proof of Capacity versus Outage Theorem \eqref{eqn:outage}}
\label{app:outage} We first prove the achievability of the capacity
versus outage theorem \eqref{eqn:outage}. Consider a fixed outage
probability $q\geq 0$.

{\em Encoding}: For any input distribution $P_{X^n}$, $\epsilon
> 0$, and $R<\ubar{\bs{I}}_q(\bs{X};\bs{Y})-\epsilon$,
generate the codebook by choosing
$X^n(1)$, $\cdots$, $X^n(2^{nR})$ i.i.d. according to the
distribution $P_{X^n}(x^n)$.

{\em Decoding}: Define, for $\epsilon>0$, the {\em typical set}
$\typ$ as
\[
\typ = \left\{ (x^n,y^n):\frac1ni_{X^nW^n}(x^n;y^n)\geq
\ubar{\bs{I}}_q(\bs{X};\bs{Y})-\epsilon \right\}.
\]
For any channel output $Y^n$, we decode as follows:
\begin{enumerate}
\item If $(X^n(i),Y^n)\not\in\typ$ for all $i\in\{1,\cdots,2^{nR}\}$,
    declare an outage;
\item Otherwise, decode to the unique index $i\in\{1,\cdots,2^{nR}\}$
such that $(X^n(i),Y^n)\in\typ$. An error is declared if more than
one such index exists.
\end{enumerate}

{\em Outage and Error Analysis}: We recall the definition of events
$E_{ji}$ in \eqref{eqn:EventEij} as
\[
E_{ji} = \left\{ \left.
(X^n(j),Y^n) \in \typ  \right| X^n(i) \,\, \text{sent} \right\}.
\]
Assuming equiprobable inputs, the expected probability of an outage
using the above scheme is:
\begin{eqnarray*}
P_o^{(n)}
    & = & \Pr \left\{ \mbox{outage}|X^n(1) \mbox{ sent} \right\} \\
    & = & \Pr \left\{ \cap_{i=1}^{2^{nR}}E_{i1}^c \right\} \\
    & \leq & \Pr \left\{ E_{11}^c \right\} \\
    & = & P_{X^nW^n}\left\{ \frac1ni_{X^nW^n}(X^n(1);Y^n)<
    \ubar{\bs{I}}_q(\bs{X};\bs{Y})-\epsilon \right\} \\
    & \leq & q + \epsilon_n,
\end{eqnarray*}
where by definition of $\ubar{\bs{I}}_q(\bs{X};\bs{Y})$ we have
$\epsilon_n$ approaching $0$ for $n$ large enough. Likewise, when no
outage is declared the expected probability of error is
\begin{eqnarray}
P_e^{(n)}
    & = & \Pr \left\{ \mbox{error}|X^n(1) \mbox{ sent and no outage declared} \right\} \nonumber \\
    & = & \Pr \left\{ \bigcup_{i=2}^{2^{nR}}E_{i1} \right\} \nonumber \\
    & \leq & 2^{nR} \Pr \left\{ E_{21} \right\} \nonumber \\
    & = & 2^{nR}\sum_{(x^n,y^n)\in\typ}P_{X^n}(x^n)P_{Y^n}(y^n) \nonumber \\
    & \leq & 2^{n[R-\ubar{\bs{I}}_q(\bs{X};\bs{Y})+\epsilon]}
    \sum_{(x^n,y^n)\in\typ}P_{X^nW^n}(x^n,y^n),
    \,\,\,\,\,\,\,\,\,\,\,\,
\label{eqn:outageBound}
\end{eqnarray}
where the last inequality is obtained by noticing that
$(x^n,y^n)\in\typ$ implies
\[
\frac1n i_{X^nW^n}(x^n;y^n) = \frac1n
\log\frac{P_{X^nW^n}(x^n,y^n)}{P_{X^n}(x^n)P_{Y^n}(y^n)} \ge
\ubar{\bs{I}}_q(\bs{X};\bs{Y})-\epsilon
\]
or equivalently
\[
P_{X^n}(x^n)P_{Y^n}(y^n)  \le
2^{-n[\ubar{\bs{I}}_q(\bs{X};\bs{Y})-\epsilon]}P_{X^nW^n}(x^n, y^n).
\]
From
\eqref{eqn:outageBound} we see that $P_e^{(n)}\to 0$ for all $R <
\ubar{\bs{I}}_q(\bs{X};\bs{Y})-\epsilon$ and arbitrary $\epsilon>0$,
which completes our proof.

Next we prove the converse of the capacity versus outage theorem
\eqref{eqn:outage}. Consider any sequence of $(n,2^{nR})$ codes with
error probability $P_e^{(n)} \to 0$ and outage probability
${\displaystyle \lim_{n\to \infty}P_o^{(n)}\leq q }$. Let
$\{X^n(1),\cdots,X^n(2^{nR})\}$ represent the $n$th code in the
sequence, and assume a uniform input distribution
\[
P_{X^n}(x^n)=\left\{
    \begin{array}{ll}
        {2^{-nR},} &\forall \,\, x^n\in \{X^n(1),\cdots,X^n(2^{nR})\}, \\
        0, & \mbox{otherwise.}
    \end{array}\right.
\]
For each $i\in\{1,\cdots,2^{nR}\}$, let $D_i$ represent the decoding
region associated with codeword $X^n(i)$ and $B_i$ equal an analogy
of the typical set, defined as
\begin{eqnarray*}
B_i & = & \left\{y^n\in {\cal Y}^n:\frac1ni_{X^nW^n}(X^n(i),y^n)\leq R-\gamma
\right \}\\
    & = & \left\{y^n\in{\cal Y}^n:
        \frac1n\log\frac{P_{X^n|Y^n}(X^n(i)|y^n)}{2^{-nR}}
        \leq R-\gamma \right\}\\
    & = & \{y^n\in{\cal Y}^n:
        P_{X^n|Y^n}(X^n(i)|y^n)\leq 2^{-\gamma n}\},
\end{eqnarray*}
where $\gamma>0$ is arbitrary. Then we have
\begin{eqnarray*}
&& P_{X^nW^n} \left\{ \frac1ni_{X^nW^n}(X^n;Y^n)\leq R-\gamma
\right\} \\
    & = & \sum_{i=1}^{2^{nR}}P_{X^nW^n}(X^n(i),B_i) \\
    & = & \sum_{i=1}^{2^{nR}}[P_{X^nW^n}(X^n(i),B_i\cap D_i) \\
    &&  +P_{X^nW^n}(X^n(i),B_i\cap D_i^c)] \\
    &\leq&\sum_{i=1}^{2^{nR}}
        \sum_{y^n\in B_i\cap D_i}P_{X^nW^n}(X^n(i),y^n)
            +P_e^{(n)}+P_o^{(n)} \\
    &\leq&\sum_{i=1}^{2^{nR}}
        \sum_{y^n\in D_i}P_{Y^n}(y^n)2^{-\gamma n}
            +P_e^{(n)}+P_o^{(n)} \\
    &\leq&2^{-\gamma n}+P_e^{(n)}+P_o^{(n)},
\end{eqnarray*}
since the decoding regions $D_i$ cannot overlap. Thus
\[
P_e^{(n)}\geq
    P_{X^nW^n} \left\{ \frac1ni_{X^nW^n}(X^n;Y^n)\leq R-\gamma
    \right\}
    -P_o^{(n)}-2^{-\gamma n},
\]
which goes to zero if and only if $R-\gamma \leq
\ubar{\bs{I}}_q(\bs{X};\bs{Y})$ by definition of
$\ubar{\bs{I}}_q(\bs{X};\bs{Y})$.

\section{Proof of Theorem \ref{thm:CeBCmapping}}
\label{app:mapping}
\subsection{Mapping Broadcast Code to Expected-rate Code}
\label{sec:BC2E} We first show that any broadcast code can be mapped
to an expected-rate code, so
\begin{equation}
C^e \ge \sum_{p \in \mathcal{P}} R_p \sum_{s\in p} P_S(s)
\label{eqn:BC2E}
\end{equation}
for any $\{R_p\} \in \mathcal{C}_{\BC}$.

Given a $(\{2^{nR_p}\}, n)$ BC code as defined in Definition
\ref{defn:BCdefn}, we represent each message $M_p \in \mathcal{M}_p$
in a binary format consisting of $nR_p$ bits and concatenate these
bits to form an overall representation of $nR_t$ bits, where
\begin{equation}
R_t = \sum_{p \in \mathcal{P}, p \neq \phi} R_p. \label{eqn:BC2ERt}
\end{equation}
These $nR_t$ information bits are indexed by the index set
$\mathcal{I}_{n,t} = \{1, 2, \cdots, nR_t\}$. We denote by
$\mathcal{I}_{n,p}$ the set of indices of the $nR_p$ bits that
correspond to the message set $\mathcal{M}_p$ in the BC code. Note
that $\mathcal{I}_{n,p}$ may be empty for some $p \in \mathcal{P}$,
for different $p$ these index sets are mutually exclusive and
\begin{equation}
\mathcal{I}_{n,t} = \bigcup_{p \in \mathcal{P}, p \neq \phi}
\mathcal{I}_{n,p}. \label{eqn:calInt}
\end{equation}

The $(\{2^{nR_p}\}, n)$ BC code can be mapped to the following
expected-rate code with transmit rate $R_t$ given by
\eqref{eqn:BC2ERt}. For any $M_t \in
\mathcal{M}(\mathcal{I}_{n,t})$, the bits $(b_i)$ with $i \in
\mathcal{I}_{n,p} \subseteq \mathcal{I}_{n,t}$ define a
corresponding message $M_p$ in the message set $\mathcal{M}_p$ of
the BC code. The encoder for the expected rate code satisfies
\[
f^e_n(M_t) = f^{\BC}_n \left( \prod_{p \in \mathcal{P}, p \neq \phi}
M_p \right),
\]
where the superscript $e$ and $\BC$ distinguishes the encoder of the
expected-rate code and the broadcast code. For a state $s$ in the
composite channel, the receiver decodes those information bits with
indices in the set
\begin{equation}
\mathcal{I}_{n,s} = \bigcup_{p: s\in p} \mathcal{I}_{n,p},
\label{eqn:calIns}
\end{equation}
and the decoding rate is $R_s = \sum_{p: s\in p}R_p$. For the
composite channel, the decoder output
\[
g^e_{n,s}(y^n) 
= (\hat{b}_i)_{i \in \mathcal{I}_{n,s}}
\]
is obtained by concatenating the binary representations
$(\hat{b}_i)_{i \in \mathcal{I}_{n,p}}$ of each $\hat{M}_p$, where
$s \in p$ and
\[
g^{\BC}_{n,s}(y^n) = \prod_{p:s \in p} \hat{M}_p
\]
is the decoder output of receiver $s$ in the broadcast channel.
The decoding error probability for the expected-rate code in channel
state $s$ is
\[
P_e^{(n,s)} = \Pr \{E_s\},
\]
where the error event $E_s$ for the broadcast code is defined in
\eqref{eqn:Es}. Notice that
\[
P_e^{(n,s)} = \Pr \{E_s\} \le \Pr \left\{ \cup_s E_s  \right\} =
P_e^{(n)}
\]
so the expected error probability
\[
\mathbb{E}_S P_e^{(n,S)} \le P_e^{(n)} \to 0
\]
as $n \to \infty$, according to the BC code definition. Therefore
the rate
\[
R = \mathbb{E}_S R_S = \sum_{s} P_S(s) R_s = \sum_{s} P_S(s)
\sum_{p: s \in p} R_p
\]
is an achievable expected rate and \eqref{eqn:BC2E} is proved.

\subsection{Mapping Expected-rate Code to Broadcast Code}
Next we show that for any fixed $\epsilon>0$,
\begin{equation}
C^e - \epsilon \le  \sup_{\{R_p\} \in \mathcal{C}_{\BC}} \sum_{p \in
\mathcal{P}} R_p \sum_{s\in p} P_S(s). \label{eqn:E2BC}
\end{equation}
According to the definition of the expected capacity, there exists a
sequence of $\{(2^{nR_t}, \{2^{nR_s}\}, n)\}$ codes such that
\begin{equation}
\mathbb{E}_S R_S \to R \ge C^e - \epsilon \label{eqn:CeR}
\end{equation}
and $\mathbb{E}_S P_e^{(n,S)} \to 0$. The transmitted information
bits are indexed by $\calIn{t}=\{1,2,\cdots,nR_t\}$. Since the
transmitter and the receiver agree on the index set $\calIn{s}$ of
those information bits that can be reliably decoded in each channel
state $s$, the transmitter can define, for each subset $p \in
\mathcal{P}$ of channel states, the index set $\calIn{p}$ of those
information bits decodable exclusively for channel states within
$p$, i.e.
\[
\calIn{p} = \left( \bigcap_{s \in p} \calIn{s} \right) \bigcap
\left( \bigcap_{s \notin p} \bar{\mathcal{I}}_{n,s} \right),
\]
where
\[
\bar{\mathcal{I}}_{n,s} = \left\{ i: i \in \calIn{t}, i \notin
\calIn{s} \right\}
\]
is the complement index set of $\calIn{s}$. Denote by $nR_p$ the
cardinality of $\calIn{p}$. We observe that $\calIn{p}$ are mutually
exclusive, the relationship \eqref{eqn:calInt} and
\eqref{eqn:calIns} still hold and the decoding rate satisfies $R_s =
\sum_{s \in p} R_p$.

The $\{(2^{nR_t}, \{2^{nR_s}\}, n)\}$ expected-rate code can be
mapped to the following BC code. Define the message set of the BC
code as
\[
\mathcal{M}_p = \mathcal{M}(\calIn{p})
\]
in the sense that each message $M_p \in \mathcal{M}_p$ has the
corresponding binary representation $(b_i)_{i \in \calIn{p}}$. The
encoder for the BC code satisfies
\[
f^{\BC}_n \left( \prod_{p \in \mathcal{P}, p \neq \phi} M_p \right)
= f^e_n(M_t),
\]
where $M_t = (b_i)_{i \in \calIn{t}}$ is obtained by concatenating
the binary representations of each $M_p$. When the composite channel
is in state $s$, the decoder output is
\[
g^e_{n,s}(y^n) = \hat{M}_s = (\hat{b}_i)_{i \in \mathcal{I}_{n,s}}.
\]
Since $\calIn{p} \subseteq \calIn{s}$ for any $p$ satisfying $s \in
p$, we define the decoder output for receiver $s$ in the BC to be
\[
g^{\BC}_{n,s} (y^n) = \prod_{p: s \in p} \hat{M}_p,
\]
where the binary representation $(b_i)_{i \in \calIn{p}}$ of each
$\hat{M}_p$ can be obtained by the corresponding bits in
$\hat{M}_s$.

The error event $E_s$ for receiver $s$ of the BC is defined in
\eqref{eqn:Es} with the error probability
\[
\Pr\{E_s\} = P_e^{(n,s)},
\]
and the overall error probability
\[
P_e^{(n)} = \Pr \left\{ \cup_s E_s  \right\} \le \sum_s \Pr\{E_s\} =
\sum_s P_e^{(n,s)}.
\]
By definition of the expected-rate capacity
\[
\mathbb{E}_S P_e^{(n,S)} = \sum_s P_S(s) P_e^{(n,s)} \ge \left(
\min_{s \in \mathcal{S}} P_S(s) \right) \left( \sum_s P_e^{(n,s)}
\right).
\]
Assuming each channel state $s$ occurs with strictly positive
probability, i.e. $ {\displaystyle \min_{s \in \mathcal{S}} P(s) >
0} $, then $\mathbb{E}_S P_e^{(n,S)} \to 0 $ implies
\[
P_e^{(n)} \le \sum_s P_e^{(n,s)} \to 0.
\]
Therefore the code constructed above is a valid BC code, i.e.
$\{R_p\} \in \mathcal{C}_{\BC}$, and we conclude
\begin{eqnarray}
R &=& \mathbb{E}_S R_S = \sum_s P_S(s) R_s = \sum_s P_S(s) \sum_{p:
s
\in p} R_p \nonumber \\
&\le& \sup_{\{R_p\} \in \mathcal{C}_{\BC}} \sum_{p \in \mathcal{P}}
R_p \sum_{s\in p} P_S(s) \label{eqn:R}.
\end{eqnarray}
From \eqref{eqn:CeR} and \eqref{eqn:R} we see the inequality
\eqref{eqn:E2BC} is established. Since $\epsilon$ is arbitrary,
Theorem \ref{thm:CeBCmapping} is a result of \eqref{eqn:BC2E} and
\eqref{eqn:E2BC}.

\section{Proof of \eqref{eqn:BECBCR}}
\label{app:BECBC} Consider a two-user BC where the channel to each
user is a BEC with erasure probability $\alpha_i$, $i=1,2$, i.e. the
conditional marginal distribution satisfies
\[
p(y_i|x) = \left\{
\begin{array}{ll}
1-\alpha_i, & y_i = x, \\
\alpha_i, & y_i = e.
\end{array}
\right.
\]
Assuming $\alpha_1 < \alpha_2$, we observe that the BC is
stochastically degraded since
\[
p(y_2|x) = \sum_{y_1} p(y_1|x)p'(y_2|y_1),
\]
where $p'(e|e) = 1$ and for $y_1 \neq e$
\[
p'(y_2|y_1) = \left\{
\begin{array}{ll}
\dfrac{1-\alpha_2}{1-\alpha_1}, & y_2 = y_1, \\
\dfrac{\alpha_2-\alpha_1}{1-\alpha_1}, & y_2 = e.
\end{array}
\right.
\]
Therefore the capacity region of the BEC-BC is the convex hull of
the closure of all $(R_1, R_{12})$ satisfying
\begin{eqnarray}
R_1 &\le& I(X;Y_1|U) \nonumber \\
R_{12} &\le& I(U;Y_2), \label{eqn:BECBCregion}
\end{eqnarray}
for some joint distribution $p(u)p(x|u)p(y_1,y_2|x)$. Since the
cardinality of the random variable $U$ is bounded by $|\mathcal{U}|
\le \min \{ |\mathcal{X}|, |\mathcal{Y}_1|, |\mathcal{Y}_2| \} = 2$
\cite[p. 422]{CoverIT} and the channel is symmetric with respect to
the alphabet $0$ and $1$, we can take $p(u)$$\sim$Bernoulli$(1/2)$
and $p(x|u)$ as the transition probability of a binary symmetric
channel with crossover probability $p$. This stochastically degraded
BEC-BC together with the auxiliary
random variable $U$ is illustrated in \refF{BECBC}. \\
\psfrag{0}{$0$}
\psfrag{1}{$1$}
\psfrag{p}{$p$}
\psfrag{e}{$e$}
\psfrag{U}{$U$}
\psfrag{X}{$X$}
\psfrag{Y1}{$Y_1$}
\psfrag{Y2}{$Y_2$}
\psfrag{1-p}{$1-p$}
\psfrag{1-alpha1}{$1-\alpha_1$}
\psfrag{alpha1}{$\alpha_1$}
\psfrag{t}{$\frac{\alpha_2-\alpha_1}{1-\alpha_1}$}
\psfrag{1-t}{$\frac{1-\alpha_2}{1-\alpha_1}$}
\begin{figure}[htbp]
\begin{center}
\includegraphics[width=3in]{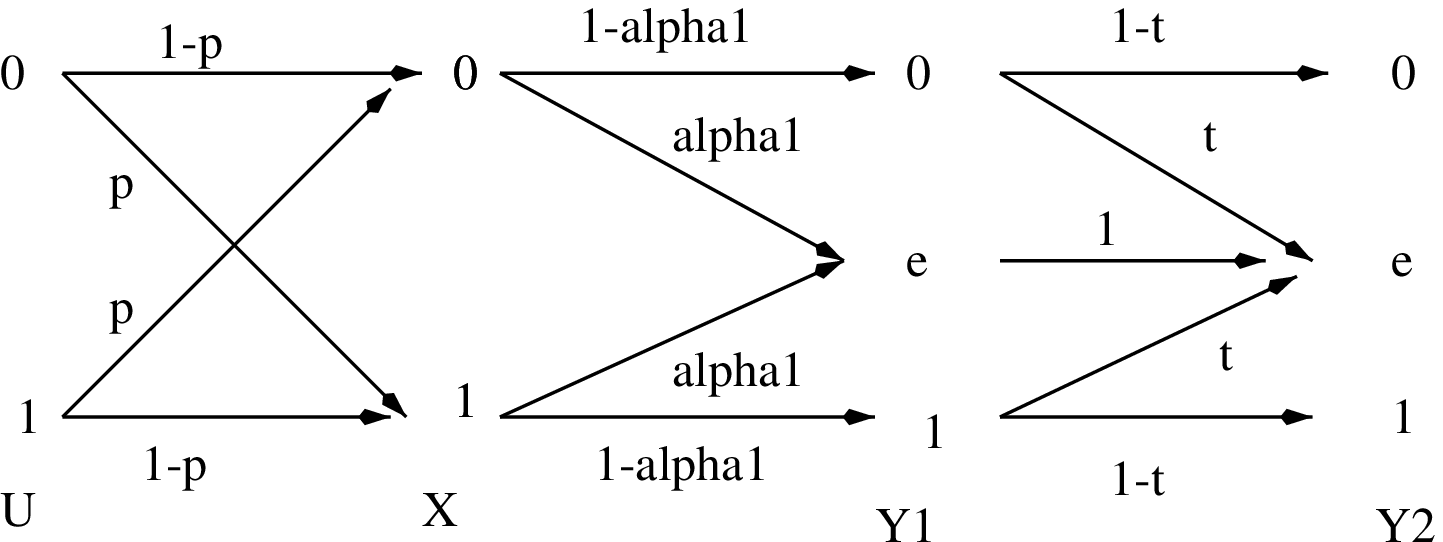}
\caption{Degraded binary erasure broadcast channel}
\label{fig:BECBC}
\end{center}
\end{figure}

The capacity region \eqref{eqn:BECBCregion} is evaluated to be
\begin{eqnarray}
R_1 &\le& (1-\alpha_1) h(p) \nonumber \\
R_{12} &\le& (1-\alpha_2) [1-h(p)], \label{eqn:BECBCregionEvaluate}
\end{eqnarray}
where $h(p)=-p\log p - (1-p) \log(1-p)$ is the binary entropy
function. Assuming the two ergodic components are equally probable
in the composite channel, the achievable expected rate using a
broadcast code is then
\begin{eqnarray*}
R &=& \sup_p \left\{ R_{12} + R_1/2 \right\} \\
&=& \max \left\{
1-\alpha_2,
\dfrac{1-\alpha_1}{2}
\right\}.
\end{eqnarray*}

\section{Proof of Upper Bound for Expected Capacity}
\label{app:upperbound}
 Denote by
$X^n_s(1)$, $\cdots$, $X^n_s(2^{nR_s})$ and $D_s(1)$, $\cdots$,
$D_s(2^{nR_s})$ the set of codewords and decoding regions
corresponding to channel
$s$. 
We fix $\gamma>0$ and define for each $s \in \mathcal{S}$ and $1 \le
i \le 2^{nR_s}$
\begin{eqnarray}
B_s(i)  & = & \{Y^n\in{\cal Y}^n:\frac1ni_{X^nW^n}(X^n(i); Y^n|s)
         \leq R_s-\gamma\} \nonumber \\
    & = & \{Y^n\in{\cal Y}^n:P_{X^n|Y^n,S}(X^n(i)|Y^n,s)
        \leq 2^{-n \gamma} \} \quad \,\, \label{eqn:Bs}
\end{eqnarray}
where \eqref{eqn:Bs} follows from \eqref{eqn:infoden}. Notice that
for any $s$ with $R_s>0$
\begin{eqnarray}
&& P_{X^nY^n|S}\left[ \left. \frac1ni_{X^nW^n}(X^n;Y^n|s)
        \leq R_s-\gamma \right|s \right] \nonumber \\
    & \leq & \sum_{i=1}^{2^{nR_s}}
   \Big[ 2^{-nR_s} P_{Y^n|X^n,S}(D_s(i)^c|X^n(i),s) \nonumber \\
&& +\sum_{y^n\in B_s(i)\cap D_s(i)}P_{X^nY^n|S}(X^n(i),y^n|s) \Big]
\nonumber \\
    & \leq & P_e^{(n,s)}+\sum_{i=1}^{2^{nR_s}}
    \sum_{y^n\in B_s(i)\cap D_s(i)} 2^{-n \gamma} P_{Y^n|S}(y^n|s)
        \nonumber \\
    & \leq & P_e^{(n,s)}+2^{-n \gamma}.
    \label{eqn:Pens}
\end{eqnarray}
Furthermore we have
\begin{eqnarray}
&& \mathbb{E}_S\liminf_{n\rightarrow\infty}P_{X^nY^n|S} \left[
\left. \frac1ni_{X^nW^n}(X^n;Y^n|S)\leq R_S-\gamma \right|S\right]
\nonumber \\
&\leq& \lim_{n\to \infty}\mathbb{E}_SP_{X^nY^n|S} \left[ \left.
\frac1ni_{X^nW^n}(X^n;Y^n|S)\leq R_S-\gamma \right|S\right]
\nonumber \\
&\leq& \lim_{n \to \infty} [ \mathbb{E}_S P_e^{(n,S)} + 2^{-n \gamma
} ] = 0, \nonumber
\end{eqnarray}
where the chain of inequalities follows from Fatou's lemma,
\eqref{eqn:Pens}, and the code constraint $\mathbb{E}_S P_e^{(n,S)}
\to 0$. Since the probability must be non-negative, we conclude
\[
\liminf_{n\rightarrow\infty}P_{X^nY^n|S}
    \left[ \left. \frac1ni_{X^nW^n}(X^n;Y^n|S)\leq R_S-\gamma \right|S \right]=0
\]
almost surely (a.s.) in $S$. Thus for any $\epsilon>0$,
\[
P_{X^nY^n|S}\left[ \left. \frac1ni_{X^nW^n}(X^n;Y^n|S)\leq
R_S-\gamma \right|S\right]<\epsilon
\]
occurs infinitely often a.s. Assuming
$\left|i_{X^nW^n}(X^n;Y^n|S)\right|$ is bounded by $M$, we then have
\[
\mathbb{E}_{X^nY^n|S}\left[ \left. \frac1ni_{X^nW^n}(X^n;Y^n|S)
\right|S \right]>(R_S-\gamma)(1-\epsilon)- \epsilon M
\]
also occurs infinitely often a.s. Since $\epsilon$ is arbitrary, we
see that
\[
\mathbb{E}_S \mathbb{E}_{X^nY^n|S}\left[ \left.
\frac1ni_{X^nW^n}(X^n;Y^n|S) \right|S \right]
\geq \mathbb{E}_SR_S-\gamma
\]
occurs infinitely often for arbitrary $\gamma$, which gives us the
upper bound \eqref{eqn:ce} for expected capacity. Note that the
expectation in the upper bound \eqref{eqn:ce} is indeed
$\frac{1}{n}I(X^n;Y^n|S)$, so the upper bound can also be proved
using the standard technique of Fano's inequality.

\bibliographystyle{IEEEtran}
\bibliography{cellular}

\end{document}